\documentclass[review]{elsarticle}

\usepackage{geometry}                
\usepackage{graphicx}
\usepackage{amssymb}
\usepackage{amsmath}
\usepackage{bm}
\usepackage{epstopdf}
\usepackage{hyperref}
\usepackage{color}
\DeclareMathAlphabet{\mathsfit}{T1}{\sfdefault}{\mddefault}{\sldefault}
\SetMathAlphabet{\mathsfit}{bold}{T1}{\sfdefault}{\bfdefault}{\sldefault}

\newcommand{\ddt}[1]{\frac{\mathrm{d} #1}{\mathrm{d}t}}
\newcommand{\ddp}[2]{\frac{\partial #1}{\partial #2}}

\newcommand{\indic}{\chi}
\newcommand{\x}{\boldsymbol{x}}
\newcommand{\y}{\boldsymbol{y}}
\newcommand{\X}{\boldsymbol{X}}
\newcommand{\heavi}{H}
\newcommand{\dirac}{\delta}
\newcommand{\R}{\mathbb{R}}
\newcommand{\grad}{\nabla}
\renewcommand{\div}{\nabla\cdot}
\newcommand{\curl}{\nabla\times}
\newcommand{\lap}{\nabla^{2}}
\newcommand{\nrm}{\boldsymbol{n}}
\newcommand{\dvol}{\mathrm{d}\x}
\newcommand{\mask}[1]{\overline{#1}}
\newcommand{\scafunc}{f}
\newcommand{\vecfunc}{\boldsymbol{u}}
\newcommand{\tensfunc}{\boldsymbol{\sigma}}

\newcommand{\surfc}{\boldsymbol{\xi}}
\newcommand{\surfpt}{\X}
\newcommand{\surfvel}{\dot{\X}}
\newcommand{\fluidvel}{\boldsymbol{v}}
\newcommand{\dsurf}{\mathrm{d}S(\surfc)}

\newcommand{\fluxvec}{\boldsymbol{Q}}
\newcommand{\sforcevec}{\boldsymbol{\sigma}}
\newcommand{\sforcescalar}{\sigma}

\newcommand{\stress}{\boldsymbol{S}}
\newcommand{\press}{p}
\newcommand{\vort}{\boldsymbol{\omega}}
\newcommand{\tot}{\mathrm{t}}
\newcommand{\vorttot}{\mask{\boldsymbol{\omega}}_{\tot}}
\newcommand{\stream}{\boldsymbol{\psi}}
\newcommand{\spot}{\varphi}

\newcommand{\gridvar}[1]{\mathsf{#1}}

\newcommand{\pointtype}[1]{\mathsfit{#1}}

\newcommand{\reg}{\gridvar{\hat{R}}}
\newcommand{\interp}{\gridvar{R}^{T}}

\newcommand{\gradgrid}{\gridvar{G}}
\newcommand{\divgrid}{\gridvar{D}}
\newcommand{\curlgrid}{\gridvar{C}}
\newcommand{\rotgrid}{\gridvar{C}^{T}}
\newcommand{\lapgrid}{\gridvar{L}}
\newcommand{\interpgrid}[2]{{}^{#2}\gridvar{I}_{#1}}
\newcommand{\lgfcomp}{g}
\newcommand{\lgf}{\gridvar{\lgfcomp}}
\newcommand{\invlapgrid}{\gridvar{L}^{-1}}
\newcommand{\idex}{\boldsymbol{i}}
\newcommand{\ddf}{\delta_{\dx}}

\newcommand{\pindex}{p}
\newcommand{\numpts}{N}
\newcommand{\gspace}[1]{\mathcal{#1}}
\newcommand{\faces}{\gspace{F}}
\newcommand{\centers}{\gspace{C}}
\newcommand{\edges}{\gspace{E}}
\newcommand{\vertices}{\gspace{V}}
\newcommand{\tensors}{\gspace{D}}
\newcommand{\unitgrid}{\gridvar{e}}
\newcommand{\onesgrid}{\gridvar{1}}
\newcommand{\gridcomp}[2]{\gridvar{#1}(#2)}
\newcommand{\vgridcomp}[3]{\gridvar{#1^{(#3)}}(#2)}
\newcommand{\ip}[2]{{\langle #1,#2 \rangle}}
\newcommand{\ipbig}[2]{{\biggl< #1,#2 \biggr>}}
\newcommand{\ipcenters}[2]{\ip{#1}{#2}_{\centers}}
\newcommand{\ipfaces}[2]{\ip{#1}{#2}_{\faces}}

\newcommand{\normfaces}[1]{||{#1}||_{\faces}}

\newcommand{\origin}{0}
\newcommand{\xcenter}[1]{\x^{\centers}(#1)}
\newcommand{\xface}[1]{\x^{\faces}(#1)}
\newcommand{\xxface}[2]{x^{\faces_{#1}}(#2)}
\newcommand{\xyface}[2]{y^{\faces_{#1}}(#2)}
\newcommand{\xedge}[1]{\x^{\edges}(#1)}

\newcommand{\spoints}{\gspace{S}^{\numpts}}
\newcommand{\vpoints}{\gspace{V}^{\numpts}}
\newcommand{\tpoints}{\gspace{T}^{\numpts}}
\newcommand{\spoint}[1]{\pointtype{#1}}
\newcommand{\vpoint}[1]{\boldsymbol{\pointtype{#1}}}
\newcommand{\tpoint}[1]{\boldsymbol{\pointtype{#1}}}
\newcommand{\onesspoint}{\spoint{1}}

\newcommand{\regds}{\gridvar{R}}
\newcommand{\heavigrid}{\gridvar{H}}
\newcommand{\hgridcenters}{\heavigrid_{\centers}}
\newcommand{\hgridfaces}{\heavigrid_{\faces}}

\newcommand{\dx}{\Delta x}
\newcommand{\dS}{\delta s}
\newcommand{\dSvec}{\delta\spoint{s}}
\newcommand{\normvec}{\vpoint{n}}
\newcommand{\xpointvec}{\vpoint{r}}
\newcommand{\had}{\circ}
\newcommand{\tensprod}{\otimes}
\newcommand{\ipscalar}[2]{\ip{#1}{#2}_{\spoints}}
\newcommand{\ipvector}[2]{\ip{#1}{#2}_{\vpoints}}
\newcommand{\ipvectorbig}[2]{\ipbig{#1}{#2}_{\vpoints}}
\newcommand{\normscalar}[1]{||#1||_{\spoints}}

\newcommand{\rhsvec}{q}
\newcommand{\solnvec}{f}
\newcommand{\diffvec}{d}
\newcommand{\scavec}{q}
\newcommand{\vecvec}{u}
\newcommand{\velvec}{v}
\newcommand{\vortvec}{w}
\newcommand{\sfvec}{s}
\newcommand{\potvec}{f}
\newcommand{\slstrength}{s}
\newcommand{\dlstrength}{d}
\newcommand{\vslstrength}{t}
\newcommand{\vdlstrength}{d}

\newcommand{\gridvorttot}{\mask{\gridvar{\vortvec}}_{\tot}}

\newcommand{\lindiff}[2]{\mathcal{L}^{#2}_{#1}}

\begin{document}
\begin{frontmatter}
\title{A method of immersed layers on Cartesian grids,\\ with application to incompressible flows}
\author{Jeff D. Eldredge\fnref{myfootnote}}
\address{Mechanical and Aerospace Engineering, 420 Westwood Plaza, University of California, Los Angeles\\ Los Angeles, CA, USA, 90095}


\fntext[myfootnote]{jdeldre@ucla.edu}

\begin{abstract}
The immersed boundary method (IBM) of Peskin (J. Comput. Phys., 1977), and derived forms such as the projection method of Taira and Colonius (J. Comput. Phys., 2007), have been useful for simulating flow physics in problems with moving interfaces on stationary grids. However, in their interface treatment, these methods do not distinguish one side from the other, but rather, apply the motion constraint to both sides, and the associated interface force is an inseparable mix of contributions from each side. In this work, we define a discrete Heaviside function, a natural companion to the familiar discrete Dirac delta function (DDF), to define a masked version of each field on the grid which, to within the error of the DDF, takes the intended value of the field on the respective sides of the interface. From this foundation we develop discrete operators and identities that are uniformly applicable to any surface geometry. We use these to develop extended forms of prototypical partial differential equations, including Poisson, convection-diffusion, and incompressible Navier-Stokes, that govern the discrete masked fields. These equations contain the familiar forcing term of the IBM, but also additional terms that regularize the jumps in field quantities onto the grid and enable us to individually specify the constraints on field behavior on each side of the interface. Drawing the connection between these terms and the layer potentials in elliptic problems, we refer to them generically as immersed layers. We demonstrate the application of the method to several representative problems, including two-dimensional incompressible flows inside a rotating cylinder and external to a rotating square.
\end{abstract}


\end{frontmatter}

\section{Introduction}
\label{sec:intro}

Over the last few decades there have been a number of computational methods proposed which discretize partial differential equations on grids that do not conform to the shape of a physical interface or boundary in the problem (e.g., a fluid--body interface).  This approach has obvious advantages for solving problems in which the interface is moving, since it replaces the need for a complex automated remeshing task with an easier one that immerses or embeds the interface into the fixed grid. These methods are often referred to generically as immersed boundary methods, though they are often classified into subclasses depending on their approach or on the authors' own identification: e.g., immersed boundary methods \cite{peskin:1j,mcqueen:1j,laipeskin:1j,roma99,fadlun:1j,kim2001immersed,taira2007,coloniustaira08:1j,liskacolonius17}; immersed interface methods \cite{levli94:1j,leelev:1j,gillis2018fast}; ghost fluid methods \cite{fedkiw1999non,liu2000boundary}; and sharp interface, cut-cell, and cut-stencil methods \cite{uday:1j,balarasyang:1j,mittal2008versatile,duan2010high}. Rather than present a comprehensive list of such methods, we refer the reader to previous reviews, such as \cite{mittalarfm:1j}.

In the immersed boundary method of Peskin \cite{peskin:1j}, which we will refer to generically in this paper as the {\em immersed boundary method} (IBM), structures of lower dimension (e.g., fibers or a membranous surface) are introduced into a higher-dimensional fluid region by applying the force they exert on the fluid via singular Dirac forcing continuously distributed over the structure. In the discrete representation of this immersion, the higher-dimensional space is discretized by a Cartesian grid and the structure is sampled by a finite number of points. The singular forcing integral is replaced by a summation over the points, with the Dirac delta function approximated by a discrete analog, designed to satisfy a number of properties on its behavior \cite{roma99}. This discrete Dirac delta function (DDF), generally constructed from a Cartesian product of one-dimensional approximations of $\delta(x)$, regularizes the force of each immersed point onto the grid.

The original immersed boundary method was designed for fluid--structure interaction: the immersed structure's motion is determined by the local fluid velocity, interpolated onto the structure's sampling points by the same DDF used for regularizing the force onto the fluid grid. This force arises from the response of the structure's dynamics to the imposed motion. Though this method works very well for thin elastic structures, it is not naturally suited for problems in which the motion of the structure is constrained in some manner, e.g., prescribed independently and/or forced to remain a rigid structure. Taira and Colonius \cite{taira2007,coloniustaira08:1j} extended the immersed boundary method to prescribed motions by interpreting the singular forcing term as distributed Lagrange multipliers (or \emph{constraint forces}) for enforcing the no-slip condition between the fluid and the structure's constrained motion. The resulting discretized system of equations takes the saddle-point form typical of constrained dynamics problems (e.g., in robotics, or in the divergence-free velocity constraint of incompressible flow, in which pressure serves as the Lagrange multiplier field). By formally solving this saddle-point system with block-LU decomposition, Taira and Colonius devised an elegant algorithm, which they named the immersed boundary projection method (IBPM), for projecting the fluid velocity field onto the set of solutions that satisfy the no-slip condition (to within the error of the interpolation provided by the DDF). Liska and Colonius \cite{liskacolonius17} later paired the method with the lattice Green's function to ensure efficient treatment of external flows.

One of the most useful aspects of these immersed boundary methods is that they require minimal geometric information about the structure itself. They require only that the structure's interface with the fluid be sampled by a set of points whose separations are approximately equal to the underlying grid spacing. In particular, the IBM and IBPM require no description of the interface orientation. As a result, the methods are relatively easy to implement compared to other methods in the larger class \cite{mittal2008versatile}. Furthermore, the solution for the constraint forces in the IBPM is significantly more efficient than if this forcing is distributed throughout the body's interior, as in the method of Glowinski et al.~\cite{glowinski98}. This simple construction has two important consequences. First, because of the lack of orientation, the method applies the same motion constraint on both sides of the interface. Second, the force that the immersed structure applies to the fluid can be interpreted as the negative of the \emph{sum} of traction forces applied to the interface by the fluid on either side. This lack of distinction between sides of the interface is obviously reasonable for simulating infinitely thin structures surrounded on both sides by the same fluid, as in most applications of Peskin's immersed boundary method, in recent application of the IBPM by Goza and Colonius to simulate coupled dynamics with thin elastic structures \cite{goza2017strongly}, or in simulations of rigid plates (e.g, \cite{taira2009three}). It is also reasonable for some cases involving thick bodies (i.e., bodies with an interior of non-zero volume), in which the interface divides space (or grid) into {\em intentional} and {\em superfluous} regions. For example, when the body is stationary, then the enforcement of zero velocity ensures both no-slip exterior to the body and a quiescent interior. Furthermore, if the body is in pure translational motion, then a uniform surface velocity leads to a uniform interior velocity, consistent with that of a rigid body.

However, the lack of distinction between sides in the IBPM is problematic for many other common scenarios. We highlight two such scenarios: rigid-body rotation and thick deformable bodies. In each of these scenarios, the central challenge is that the constraint force, obtained as part of the overall saddle-point solution, does not solely represent the traction exerted by the flow in the intentional region, but rather, is an indistinguishable mix of tractions applied by the flows in the intentional and superfluous regions. For thick deformable bodies, one requires (but cannot obtain) the local traction from the fluid in the elasticity equations that govern the true interior region. When simulating the flow external to a rigidly-rotating body, the superfluous motion established in the interior is inconsistent with that of rigid-body rotation except at steady state. As a result, the force and moment obtained from integrating the constraint force does not predict those actually exerted by the exterior fluid. This issue arose in a pair of studies that extended the IBPM to coupled fluid/rigid-body dynamics \cite{wang2015strongly,lacis2016stable}. Wang and Eldredge \cite{wang2015strongly} partly addressed it by subtracting the force and moment expected from the dynamical equation for a rigidly-moving fluid, but omitted any treatment of the non-rigid motion. L\=acis et al.~\cite{lacis2016stable} also subtracted the interior dynamics, integrating the computed fluid motion in the superfluous region to account for this issue more accurately. It should be noted that the specific issue with rigid-body rotation of a single body can be overcome by expressing the problem in the body's coordinate system \cite{tsai2016coriolis}; however, this technique cannot be extended to multiple bodies in relative motion.

We also note that this spurious flow generation is not unique to members of the Peskin-type IBM family. Many other methods in the larger class avoid any complication from this by using techniques that preferentially distribute the prescribed surface motion toward one side, e.g. by using velocity interpolation stencils with ghost points, as in the methods of Mittal et al.~\cite{mittal2008versatile} and Balaras and Yang \cite{balarasyang:1j}, setting the velocity explicitly to zero inside the body. These methods impose the surface velocity constraint directly in the governing equations, rather than relying on Lagrange multipliers. Since they do not exploit these Lagrange multipliers for calculating surface tractions, they are not affected by the challenges of disentangling these data from each side. However, to achieve this one-sidedness, they generally require nontrivial implementations to adapt the interpolation to various surface geometries. Furthermore, the constraints remain the same in all immersed boundary methods, and the framework we present in this work can provide a common foundation on which to analyze and compare various immersed boundary approaches.

\begin{figure}[tbp]
\begin{center}
\includegraphics[width=0.35\textwidth]{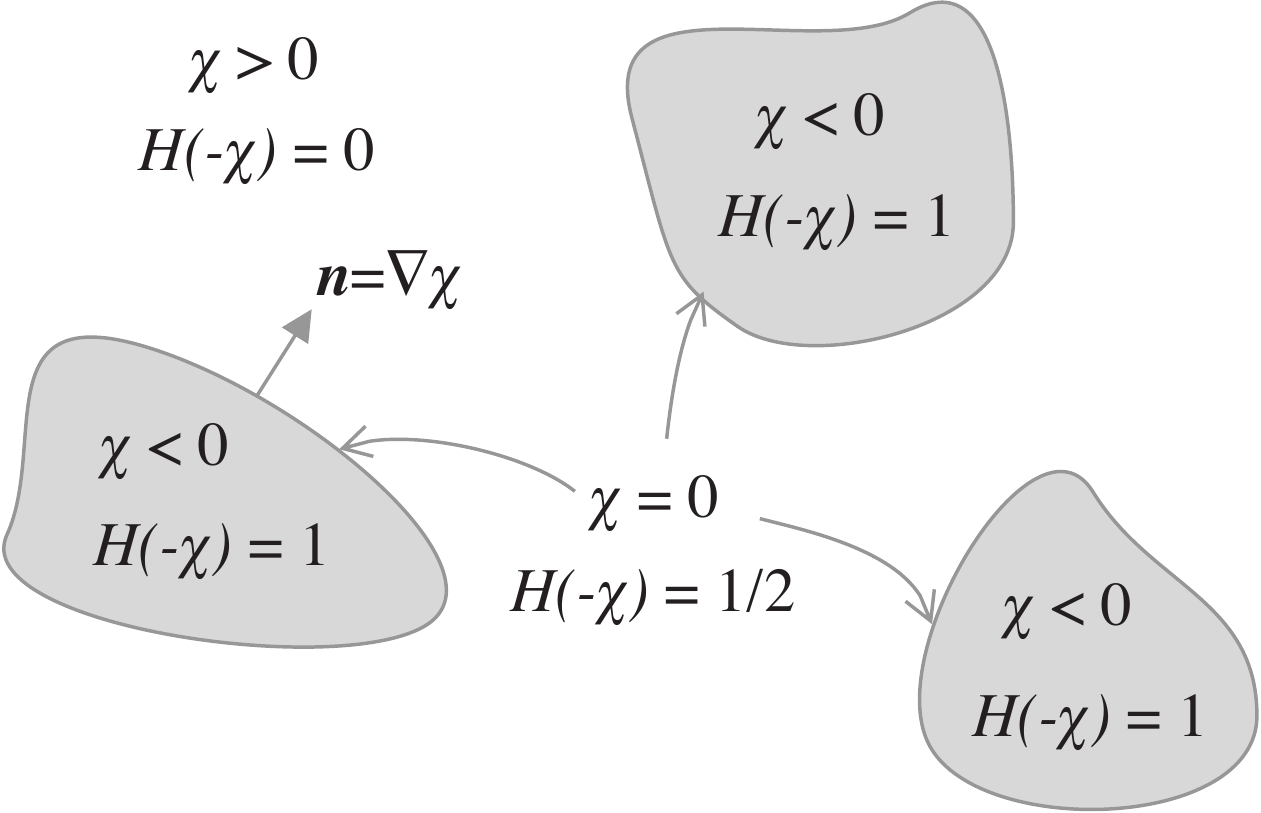}
\caption{Illustration of the indicator function, $\indic$, its level sets $\indic=0$, and the exterior ($\indic > 0$) and interior ($\indic < 0$) regions. Values of the Heaviside function $\heavi(-\indic)$ in each region are also shown; $\heavi(\indic)$ has the opposite behavior.}
\label{fig:indicatorfcn}
\end{center}
\end{figure}

In this work, we address these issues by developing a form of the immersed boundary method that distinguishes one side of an interface from the other, with simple operators that are uniformly adaptable to different surface geometries. We do this in two parts. In the first part, described in Section~\ref{sec:modelpdes}, we revisit and expand upon the original principle that leads to the singular forcing term in the continuous governing equations of the immersed boundary method. For the intermediate development, we make thorough use of the Heaviside function, $\heavi$, applied to an indicator function, $\indic$ (also known as a level set function). The composites $\heavi(\indic)$ and $\heavi(-\indic)$ represent masks that assign unity to one side of the interface and zero to the other, as illustrated in Figure~\ref{fig:indicatorfcn}. We use these to define a {\em masked} field quantity,
\begin{equation}
\label{eq:maskfone}
\mask{f} \equiv f^{+} \heavi(\indic) + f^{-} \heavi(-\indic),
\end{equation}
that extends the smooth fields $f^{+}$ and $f^{-}$ defined on each side of the interface to the full Euclidean space. Singular integrals, containing jumps in the field quantity across the interface, naturally emerge when spatial or temporal derivatives are applied to this masked field. For example, taking the gradient of the masked field, we get
\begin{equation}
\label{eq:gradf00}
\grad \mask{f} = \heavi(\indic) \grad f^{+} + \heavi(-\indic) \grad f^{-} + (f^{+}-f^{-}) \dirac(\indic) \nrm,
\end{equation}
in which the first two terms comprise a masked form of the gradient field and the final term immerses the jump on the interface into the Euclidean space with a singular integral. \ref{sec:genfun} presents a brief development of other useful identities.

With this set of identities on derivatives, the partial differential equation governing the field in either side of the interface can be reformulated as a single equation for the masked field, containing singular terms for the jumps in the field. In Section~\ref{sec:modelpdes} we carry this out for several prototypical partial differential equations: the Poisson equation, the convection-diffusion equation, and the Navier--Stokes equations. In the case of the Poisson equation, the treatment leads to a modified form of the equation expressing the familiar double- and single-layer potentials---with jumps in the field and its normal derivative---as singular forcing terms, which we refer to as {\em immersed layers}. We adopt this immersed-layer interpretation of the singular terms that emerge in the other equations, as well. In the case of the incompressible Navier--Stokes and continuity equations, we show that their extended forms for the masked velocity and pressure contain layer terms with jumps in the traction and velocity. The familiar immersed boundary method form emerges only when it is assumed that the jump in velocity is zero---that is, when the motion is imposed identically on both sides of the interface. Thus, the velocity jump terms enable us to distinguish the treatment of each side, and in particular, to force the motion on one side to zero, and thence to obtain the correct traction on the other. We also write the extended Navier--Stokes equation in vorticity form, and show that the curl of the masked velocity field combines the fluid vorticity with a vortex sheet on the interface due to the jump in velocity.

In the second part of the paper, covered in Section~\ref{sec:discretemasks}, we develop discrete forms of the immersed layers, underpinned by the same discrete Dirac delta function used in previous immersed boundary methods. Guided by the continuous equations, we build a discrete Heaviside function on this DDF and demonstrate its utility for masking data on the grid. We present the development of the discrete immersed layers in Section~\ref{sec:discretemasks} with the help of second-order mimetic operators on a staggered Cartesian grid: differencing and interpolation operators and a lattice Green's function for the discrete Laplace operator, outlined in \ref{sec:discrete}. Though the basic ideas of this paper can be carried out with other discretization methods, this treatment ensures that the immersed layers emerge naturally from discrete identities of the differencing of the masked data (and their products), each of which mimics an analogous continuous identity, such as equation~(\ref{eq:gradf00}). It should be noted that the relatively minor price we pay for our approach is that we must assign a surface normal vector to every immersed point. Indeed, the additional terms in our method generally involve the immersion of these surface normals (along with other surface data) onto the grid.

Finally, in Section~\ref{sec:soln}, we use the discrete immersed layers to develop semi-discrete forms of the prototype equations. For each equation, we pose one or more representative problems with Dirichlet boundary conditions, identify one of the immersed layer terms with the Lagrange multiplier, and solve the resulting saddle-point system. We compare the results with the exact solution where possible. We note that, with our use of indicator and Heaviside functions, the formulation we arrive at is reminiscent of some previous works that have used these tools to implicitly identify interfaces for imposing conditions. However, we have not seen any previous works that have defined a discrete Heaviside function to obtain entirely separate terms in the governing equations. (Zhao et al.~\cite{zhao2008fixed} used the Heaviside of the indicator function to distinguish elastic and viscous fluid regions, but did not fully exploit this function to incorporate jump conditions.)

We note that our treatment is still subject to the same error that affects the original immersed boundary method and the IBPM, since we do not extend the fields on either side into the opposite region. As a result, we obtain first-order convergence for the boundary error in each of our tests. It should be emphasized, however, that our goal in this work is not to improve upon this error---which can be pursued in future work---but to preserve the ease of formulation and implementation of the IBPM and extend its range of applicability. We demonstrate this extension by solving for the flow in the interior of a rotating cylinder and predicting the required moment on the cylinder, a prediction that cannot be readily made by the IBPM. We also demonstrate the method on the flow generated external to a rotating square.


\section{Prototype partial differential equations with immersed layers}
\label{sec:modelpdes}

In the introduction, we briefly discussed the masking functions $\heavi(\indic)$ and $\heavi(-\indic)$ which assign unity (resp., zero) and zero (resp., unity) to regions $\indic>0$ and $\indic<0$, and $1/2$ to the level set $\indic=0$. In \ref{sec:genfun} we expand on the properties of these generalized functions, and define an associated {\em immersion function}, $\dirac(\indic)$, from their gradient,
\begin{equation}
\label{eq:gradheavi0}
\grad\heavi(\pm\indic) =\pm \dirac(\indic) \nrm,
\end{equation}
where $\nrm$ is the unit normal as defined in Figure~\ref{fig:indicatorfcn}. This function, defined in (\ref{eq:diracg3}), immerses surface data on $\indic=0$ into Euclidean space via a singular integral; it has a companion {\em restriction function}, $\dirac^{T}(\indic)$, which restricts data in the full space onto the surface $\indic=0$. We review several of the properties of these generalized functions in \ref{sec:genfun}, and importantly, demonstrate their use in spatial and temporal derivatives of a masked field, $\mask{f}$, defined as in (\ref{eq:maskfone}); we have already seen the example of the gradient operation in (\ref{eq:gradf00}). In this section, we use these properties and identities to derive versions of several model governing equations for masked fields that hold in the full space $\R^{d}$ ($d = 2$ or $3$) when this space is divided into exterior and interior regions defined by $\indic = 0$. 

\subsection{The Poisson equation and its solution}
\label{sec:poisson}

Let us suppose that $\varphi$ is a piecewise twice-differentiable scalar field in each region separated by $\indic = 0$, vanishing at infinity. We can obviously define the masked form $\mask{\varphi}$ in the same manner as (\ref{eq:maskfone}). The gradient of $\mask{\varphi}$ follows from (\ref{eq:gradf00}). And furthermore, the divergence of the gradient follows from (\ref{eq:divf}). From these, it is easy to show from the identities in \ref{sec:genfun} that
\begin{equation}
\label{eq:lapphi}
\lap \mask{\varphi} = \mask{\lap\varphi}  + (\grad\varphi^{+} - \grad\varphi^{-})\cdot\dirac(\indic)\nrm  + \div \left( (\varphi^{+}-\varphi^{-}) \dirac(\indic) \nrm \right),
\end{equation}
where, using our notation (\ref{eq:maskDf}), we have defined a masked form of the Laplacian operation, 
\begin{equation}
\mask{\lap\varphi}  = \heavi(\indic) \lap\varphi^{+} + \heavi(-\indic) \lap\varphi^{-}.
\end{equation}

Suppose that each $\varphi^{\pm}$ individually satisfies a Poisson equation in its respective region:
\begin{equation}
\lap\varphi^{+} = q^{+},\qquad \lap\varphi^{-} = q^{-},
\label{eq:poissons}
\end{equation}
where $q^{+}$ and $q^{-}$ are integrable source functions. Then $\mask{\lap\varphi} = \mask{q}$, and equation~(\ref{eq:lapphi}) with this substitution represents a generalized form of the Poisson equation for the function $\mask{\varphi}$. The Laplacian of the masked form $\mask{\varphi}$ explicitly introduces jumps on the surface $\indic=0$ in both the function itself as well as its normal derivative into an augmented form of the source function. In fact, these are familiar as single- and double-layer potentials, respectively, with corresponding strengths $\nrm\cdot(\grad\varphi^{+} - \grad\varphi^{-})$ and $\varphi^{+}-\varphi^{-}$. We can formally solve the generalized Poisson equation (\ref{eq:lapphi}) with the help of Green's function of the (negative of the) Laplacian, $G$, defined by
\begin{equation}
\label{eq:green}
\lap G(\x) = -\dirac(\x).
\end{equation}
In \ref{sec:green} we develop the Green's function solution to equation (\ref{eq:lapphi}),
\begin{align}
\label{eq:phisoln}
\mask{\varphi}(\y) &= -\int_{\R^{d}} G(\x-\y) \mask{q}(\x) \,\dvol  - \int_{\R^{d}}  G(\x-\y) \nrm\cdot(\grad\varphi^{+} - \grad\varphi^{-}) \dirac(\indic)\,\dvol \nonumber \\
&  \hspace{5cm} - \nabla_{\y} \cdot \int_{\R^{d}} G(\x - \y) (\varphi^{+}-\varphi^{-}) \dirac(\indic) \nrm\,\dvol.
\end{align}
By identity (\ref{eq:diracg}), the second and third integrals are equivalent to surface integrals on $\indic=0$, expressing the influence of the jumps in the normal derivative of $\varphi$ and of $\varphi$ itself on this surface, corresponding to standard single- and double-layer potentials, respectively. Unsurprisingly, the solution (\ref{eq:phisoln}) is equivalent to the Green's function solution of the usual Poisson equation (obtained from integrals over each region).

The Green's function form (\ref{eq:phisoln}) serves as a starting point for solving problems with known boundary data \cite{stakgold2000boundary}. For example, suppose that we seek the solution for discontinuous Dirichlet boundary conditions, $\varphi^{+} = \varphi^{+}_{b}$ and $\varphi^{-} = \varphi^{-}_{b}$ on the surface $\indic=0$. We would substitute the jump in these known values into the double-layer potential, and seek the unknown jump in normal derivatives in the single-layer potential by setting the full expression (\ref{eq:phisoln}) equal to the known value of $\mask{\varphi}$ on the surface, $(\varphi^{+}_{b} + \varphi^{-}_{b})/2$. However, for our later use in the discrete problem, it is helpful to express this procedure in a saddle-point form. We denote the unknown strength distribution of the single-layer potential by $\sforcescalar = \nrm\cdot(\grad\varphi^{+} - \grad\varphi^{-})$. Then, we can write the problem for $\mask{\varphi}$ and $\sforcescalar$ as
\begin{equation}
\label{eq:Dirichlet-matrix}
\begin{bmatrix}
\lap & \dirac(\indic) \\ \dirac^{T}(\indic) & 0
\end{bmatrix}
\begin{pmatrix}
\mask{\varphi} \\ -\sforcescalar
\end{pmatrix} = 
\begin{pmatrix}
\mask{q}   + \div \left( (\varphi^{+}_{b}-\varphi^{-}_{b}) \dirac(\indic) \nrm \right) \\ \frac{1}{2} \left( \varphi^{+}_{b} + \varphi^{-}_{b}\right)
\end{pmatrix} 
\end{equation}
which emphasizes the role of the single-layer strength $\sforcescalar$ as a Lagrange multiplier for enforcing the Dirichlet conditions on the masked function $\mask{\varphi}$. This form demonstrates the roles of the companion functions, $\dirac(\indic)$ and $\dirac^{T}(\indic)$ for, respectively, immersing surface data into the full space and for restricting a full-space function onto the surface. We will pose the problem in this form later in the discrete context, where we will also show that these functions become the familiar regularization and interpolation operators---transposes of one another---in immersed boundary methods.

An important special case of (\ref{eq:lapphi}) and its solution (\ref{eq:phisoln}) emerges when we take $\varphi^{+} = 0$ and $\varphi^{-} = 1$, both of which are trivial solutions of Laplace's equations in their respective regions. The masked function corresponding to these is simply $\mask{\varphi} = \heavi(-\indic)$, and equation~(\ref{eq:lapphi}) reduces to
\begin{equation}
\label{eq:lapheavi}
\lap \heavi(-\indic) = - \div \left( \dirac(\indic) \nrm \right).
\end{equation}
This equation, which can be interpreted as a governing equation for the interior masking function $\heavi(-\indic)$, will serve a crucial role in our development of discrete masking functions later in the paper. The solution (\ref{eq:phisoln}) reduces to
\begin{equation}
\label{eq:interiormask}
\heavi(-\indic(\y)) = \nabla_{\y} \cdot \int_{\R^{d}} G(\x - \y)\dirac(\indic) \nrm\,\dvol.
\end{equation}
In this manner, the interior mask is effectively generated by a double-layer potential of strength $-1$ on the level set $\indic = 0$. The exterior mask $\heavi(\indic)$ follows immediately from the fact that $\heavi(\indic) = 1 - \heavi(-\indic)$. We will develop discrete analogs of these expressions in Section~\ref{sec:discrete}.

We note, in passing, that one is tempted to use (\ref{eq:phisoln}) to form an expression similar to (\ref{eq:interiormask}) for this exterior mask. However, we cannot simply set $\varphi^{+}=1$ and $\varphi^{-}=0$, since we have assumed that $\varphi^{+}$ vanishes at infinity. Indeed, we cannot obtain the uniform value of $1$---a homogeneous solution of the Poisson equation---by means of the Green's theorem (\ref{eq:phisoln}).



\subsection{The convection-diffusion equation} Let us now consider a generic inhomogeneous convection-diffusion equation for a scalar quantity $\varphi$
\begin{equation}
\ddp{\varphi}{t} + \fluidvel\cdot\grad \varphi = -\div \fluxvec + q, 
\end{equation}  
where $\fluidvel$ is the velocity of the fluid, $q$ is a volumetric forcing term, and the flux vector $\fluxvec$ has the typical form $\fluxvec = -\kappa \grad \varphi$. We will assume that this equation governs both the interior and exterior regions. For simplicity, we will also assume that the scalar diffusion coefficient $\kappa$ is uniform and identical in both regions, though we are not restricted to do so. We wish to write this equation in masked form, with each of the quantities in the equation replaced by a masked version. Using the results obtained in \ref{sec:genindic}, it is straightforward to show that this form is
\begin{equation}
\ddp{\mask{\varphi}}{t} + \mask{\fluidvel}\cdot\grad \mask{\varphi} = -\div \mask{\fluxvec} + \mask{q} + \sforcescalar\dirac(\indic),
\label{eq:convdiff} 
\end{equation}
where we have defined the surface force function, $\sforcescalar$, as
\begin{equation}
\label{eq:sforcedef}
\sforcescalar = (\fluxvec^{+}-\fluxvec^{-})\cdot\nrm + (\varphi^{+} -\varphi^{-})(\mask{\fluidvel}-\surfvel)\cdot\nrm,
\end{equation}
where $\surfvel$ denotes the local velocity of the surface $\indic=0$. The function $\sforcescalar$ represents the net diffusive plus convective flux through the surface. Using the form of the flux vector and equation~(\ref{eq:gradf00}), we can write the masked version of this flux vector as
\begin{equation}
\label{eq:fluxvec}
\mask{\fluxvec} = -\kappa \grad \mask{\varphi} + \kappa (\varphi^{+} - \varphi^{-}) \dirac(\indic)\nrm,
\end{equation}
and the final form of the masked convection-diffusion equation is thus
\begin{equation}
\label{eq:maskedconvdiff}
\ddp{\mask{\varphi}}{t} + \mask{\fluidvel}\cdot\grad \mask{\varphi} = \kappa \lap \mask{\varphi} + \mask{q} + \sforcescalar\dirac(\indic) - \div \left[ \kappa (\varphi^{+} - \varphi^{-}) \dirac(\indic)\nrm\right].
\end{equation}

When Dirichlet conditions $\varphi^{\pm} = \varphi^{\pm}_{b}$ are imposed on the interface $\indic=0$, the final term in (\ref{eq:maskedconvdiff}) becomes a known forcing (similar to the double layer in the Poisson equation) and the surface force function $\sforcescalar$ serves, as in the case of the Poisson equation, as a Lagrange multiplier for the constraint $\dirac^{T}(\indic) \mask{\varphi} = \frac{1}{2}(\varphi^{+}_{b} + \varphi^{-}_{b})$. Once $\sforcescalar$ is found as part of the solution, it can be interpreted physically via (\ref{eq:sforcedef}).


\subsection{The incompressible Navier--Stokes equations} In this section, we formulate the Navier--Stokes equations for incompressible flow in a masked form appropriate for all of $\R^{d}$, when these equations govern the regions both interior and exterior to the interface $\indic=0$, i.e.,
\begin{equation}
\rho\ddp{\fluidvel}{t} + \rho \fluidvel\cdot\grad\fluidvel = \div \stress,\quad \stress = -\press \boldsymbol{I} + \mu \left( \grad \fluidvel + \grad^{T} \fluidvel\right), \quad \div \fluidvel = 0,
\end{equation}
where $\boldsymbol{I}$ is the identity tensor in $\R^{d}$. We will assume that the viscosity $\mu$ is uniform and identical in each region. Much of the work we performed in developing the convection-diffusion equation (\ref{eq:maskedconvdiff}) is immediately useful here, with some minor adaptations needed to accommodate the tensor form of $\stress$. Here, the equations become
\begin{align}
\label{eq:maskedns}
\rho \left(\ddp{\mask{\fluidvel}}{t} +  \mask{\fluidvel}\cdot\grad\mask{\fluidvel} \right) &= -\grad \mask{\press} + \mu \lap \mask{\fluidvel} - \sforcevec \dirac(\indic) - \div \left( \boldsymbol{\Sigma} \dirac(\indic)\right), \\
 \div \mask{\fluidvel} &= (\fluidvel^{+}-\fluidvel^{-})\cdot\dirac(\indic)\nrm. \nonumber
\end{align}
The right-hand side of the continuity equation represents a distributed set of sources that correct for a possible jump in normal velocity on the interface. The surface force function $\sforcevec$ is the sum of surface tractions on either side, plus the jump in momentum fluxes through the surface,
\begin{equation}
\label{eq:surfaceflux}
\sforcevec = (\stress^{+}-\stress^{-})\cdot \nrm - \rho (\fluidvel^{+}-\fluidvel^{-})(\mask{\fluidvel}-\surfvel)\cdot\nrm,
\end{equation}
and we have defined the symmetric viscous surface tensor $\boldsymbol{\Sigma}$ as
\begin{equation}
\label{eq:viscoussurf}
\boldsymbol{\Sigma} =  \mu \left[ (\fluidvel^{+}-\fluidvel^{-})\nrm + \nrm (\fluidvel^{+}-\fluidvel^{-}) \right].
\end{equation}
This surface tensor introduces the interface's velocity jump into the rate of deformation tensor in the Newtonian fluid model. As in the previous model equations, the surface force function $\sforcevec$ can serve as a Lagrange multiplier for enforcing the Dirichlet (no-slip) boundary condition for $\mask{\fluidvel}$ on the surface $\indic=0$,
\begin{equation}
\label{eq:noslip}
\dirac^{T}(\indic)\mask{\fluidvel} = \frac{1}{2} \left( \fluidvel^{+}_{b} + \fluidvel^{-}_{b} \right).
\end{equation}

Thus far, these equations are quite generally applicable to several scenarios. Let us suppose our goal is to solve for a flow exterior to a closed body whose bounding surface is defined by the level set $\indic=0$ and moves with local velocity $\surfvel = \fluidvel_{b}$. Thus, the boundary condition we seek to enforce on this surface is $\fluidvel_{b}^{+}=\fluidvel_{b}$. We will discuss two possible approaches to using equations (\ref{eq:maskedns}) to solve this flow, which differ in the manner in which we set a condition on $\fluidvel_{b}^{-}$ in the superfluous interior region.

In the first approach, we also set $\fluidvel_{b}^{-} = \fluidvel_{b}$, and the governing equations (\ref{eq:maskedns}) reduce to
\begin{align}
\label{eq:maskedns1}
\rho \left(\ddp{\mask{\fluidvel}}{t} + \mask{\fluidvel}\cdot\grad\mask{\fluidvel} \right) &= -\grad \mask{\press} + \mu \lap \mask{\fluidvel} - \sforcevec \dirac(\indic), \nonumber\\
\div \mask{\fluidvel} &= 0, \\
\dirac^{T}(\indic)\mask{\fluidvel} &= \fluidvel_{b} \nonumber.
\end{align}
These equations are of the form used in the immersed boundary method \cite{peskin:1j} and the immersed boundary projection method \cite{taira2007}. It is important to observe that the surface force function $\sforcevec$ is equal to $(\stress^{+}-\stress^{-})\cdot \nrm$, the sum of the surface tractions exerted from either side, in this approach. Thus, if one desires the exterior traction $\stress^{+}\cdot\nrm$, it is not possible to obtain it from $\sforcevec$ without some independent knowledge of the flow generated in the interior region. Such knowledge is available in special cases: a stationary surface, in which case $\fluidvel^{-}$ is identically zero throughout the interior region, so $\stress^{-}$ itself is zero; and uniform $\fluidvel_{b}$---i.e., rigid-body translation---so that $\fluidvel^{-} = \fluidvel_{b}$ is the solution at all times in the interior. In this latter case, the integral of the interior traction $\stress^{-}\cdot\nrm$ is equal to the rate of change of rigid-body momentum in the interior \cite{wang2015strongly}, and its local value could be determined by considering the distributed force required to ensure rigid-body motion. For cases of rigid-body rotation or deforming body motion, however, a non-trivial flow is generated in the interior by the imposed boundary condition.

In this first approach, our difficulty arises from keeping the exterior and interior regions coupled to one another via their shared boundary condition. Thus, in the second approach, we isolate the exterior region from the interior by explicitly setting the interior boundary condition to zero, $\fluidvel_{b}^{-} = 0$, ensuring that the velocity and stress throughout this region is also zero. This results in the following set of equations for $\mask{\fluidvel}$, slightly more complicated than in the first approach:
 \begin{align}
\label{eq:maskedns2}
\rho \left(\ddp{\mask{\fluidvel}}{t} +  \mask{\fluidvel}\cdot\grad\mask{\fluidvel}\right) &= -\grad \mask{\press} + \mu \lap \mask{\fluidvel} - \sforcevec \dirac(\indic) - \div \left[ \mu (\fluidvel_{b}\nrm +  \nrm \fluidvel_{b}) \dirac(\indic)\right], \nonumber\\
 \div \mask{\fluidvel} &= \fluidvel_{b}\cdot\dirac(\indic)\nrm, \\
 \dirac^{T}(\indic)\mask{\fluidvel} &= \frac{1}{2}\fluidvel_{b} \nonumber.
 \end{align}
Now, the surface traction can be easily found from the Lagrange multiplier $\sforcevec$ and the known momentum flux by a slight re-arrangement of (\ref{eq:surfaceflux}):
 \begin{equation}
 \label{eq:tractionext}
\stress^{+}\cdot \nrm =  \sforcevec - \frac{1}{2}\rho \fluidvel_{b}\fluidvel_{b}\cdot\nrm.
\end{equation}
The overall force and moment on any body can easily be obtained by integrating. A similar approach can be used, simply by setting $\fluidvel_{b}^{+} = 0$ and $\fluidvel_{b}^{-} = \fluidvel_{b}$, if the desire is to simulate an interior flow and preserve quiescent flow in the exterior region.

\subsection{Vorticity form of the incompressible Navier--Stokes equations}

It is useful to express the incompressible flow equations for $\mask{\fluidvel}$ in vorticity form. We will use the general set of equations (\ref{eq:maskedns}) as a starting point, and either of the approaches we outlined in the previous section can be expressed in a vorticity form with the equations we establish here. We first note that, from (\ref{eq:curlf}), the curl of the masked velocity has the form
\begin{equation}
\label{eq:vorttot}
\vorttot \equiv \curl \mask{\fluidvel} = \mask{\vort} + \dirac(\indic) \nrm \times (\fluidvel^{+}-\fluidvel^{-}) .
\end{equation}
The first term on the right-hand side is the masked vorticity, bounded everywhere but possibly discontinuous on the interface. The second term represents the strength of a vortex sheet, associated with the jump in velocity across the interface. For shorthand, we have denoted the overall curl of velocity, including the singular part associated with the vortex sheet, by $\vorttot$. We will refer to this as the augmented vorticity field.

The vorticity form of Navier-Stokes arises, as usual, by taking the curl of these equations. It follows easily that
\begin{equation}
\label{eq:maskednsvort}
\rho\left(\ddp{\vorttot}{t} - \curl(\mask{\fluidvel}\times\vorttot) \right) = \mu \lap \vorttot - \curl (\sforcevec \dirac(\indic)) - \curl\left[\div \left( \boldsymbol{\Sigma} \dirac(\indic)\right)\right],
\end{equation}
in which $\sforcevec$ remains the Lagrange multiplier for the boundary condition on the velocity, the no-slip condition (\ref{eq:noslip}). It should be noted that this set of equations holds for the augmented vorticity field, $\vorttot$, including the vortex sheet associated with the jump in interface velocities.

To recover the velocity from the vorticity field, we rely as usual on a vector potential, $\mask{\stream}$ (streamfunction in two dimensions), assumed to have zero divergence. The masked velocity field cannot be written entirely as the curl of a vector potential because of the jump in normal velocities on the right-hand side of the continuity equation; this calls for an additional scalar potential field, $\mask{\spot}$. We also include a uniform flow, if present, and write the overall masked velocity field using the Helmholtz decomposition as
\begin{equation}
\label{eq:helm}
\mask{\fluidvel} = \curl \mask{\stream} + \grad \mask{\spot} + \boldsymbol{U}_{\infty}.
\end{equation}
 It should be noted that, in obtaining this masked form of the decomposition, we have required that the exterior and interior values $\stream^{\pm}$ and $\spot^{\pm}$ each match across the interface to eliminate singular terms in these potentials and ensure that the velocity field is non-singular. By taking the curl and divergence, respectively, of (\ref{eq:helm}), we obtain Poisson equations for each of the potentials,
\begin{align}
\lap \mask{\stream} &= -\vorttot = -\mask{\vort} + (\fluidvel^{+}-\fluidvel^{-})\times \dirac(\indic) \nrm, \\
\lap \mask{\spot} &= (\fluidvel^{+}-\fluidvel^{-})\cdot\dirac(\indic)\nrm.
\end{align}
The only boundary conditions on $\mask{\stream}$ and $\mask{\spot}$ are that they should vanish at infinity. Each of them is readily solved by the Green's function treatment described in Section~\ref{sec:poisson}. The velocity in (\ref{eq:helm}) is constituted from the solutions. Thus, with these equations and (\ref{eq:maskednsvort}), we obtain a vorticity form of Navier--Stokes in which we can apply a specific set of conditions on the interface, as in the velocity--pressure form discussed in the previous section.

\section{Discrete masking functions and immersed single and double layers}
\label{sec:discretemasks}
 
In this section, we discuss discrete versions of the generalized functions, $\dirac(\indic)$ and $\heavi(\indic)$, and develop key relationships that will support the discretization of partial differential equations in the next section. In \ref{sec:discrete}, we define spaces on a Cartesian grid (cell centers $\centers$, faces $\faces$, and edges $\edges$), on the $\numpts$ discrete points that sample the surface(s) (with scalar $\spoints$, vector $\vpoints$, and tensor $\tpoints$ data), and discrete forms of the operators between these spaces. These operators and their associated identities mimic the continuous ones used in the previous section and detailed in \ref{sec:genfun}. We note that we will occasionally refer to an element of one of these spaces as a ``vector'' in the linear algebra sense, e.g., $\spoint{s} \in \spoints$. This terminology should not be confused with the type of data it assigns to each point, which is manifested by the name of the space (e.g., scalar-valued data in $\spoints$). Elements in the grid spaces will be written in roman sans-serif font, e.g., $\gridvar{w} \in \edges$, while those in the immersed point spaces will be shown in italicized sans-serif, e.g., $\spoint{s} \in \spoints$ or $\vpoint{v} \in \vpoints$.

It should be emphasized that the core ideas of the framework we introduce in this paper can be applied on any structured grid with finite difference methods or on unstructured meshes with finite volume methods. To be definite in this paper, however, we assume the fields are expressed on a staggered Cartesian grid of uniform spacing $\dx$ and infinite extent, and second-order differencing and interpolation operators on this grid. In practice, of course, the grid can be finite, but our treatment does not require any conditions on the boundary of the grid, so we can treat it as notionally infinite for our derivations. With the help of a lattice Green's function, the discrete Laplacian $\lapgrid$ is paired with an inverse operator, $\invlapgrid$, in equation~(\ref{eq:lapgrid}). The immersion process is based on a discrete Dirac delta function (DDF), $\ddf(\x)$, which serves as an approximation of $\dirac(\x)$ \cite{peskin:1j,roma99,taira2007}. This DDF leads to conventional definitions of the regularization $\regds$ and interpolation $\interp$ operators \cite{roma99}---the discrete versions of immersion $\dirac(\indic)$ and restriction $\dirac^{T}(\indic)$. Because these definitions are common, we relegate the details to \ref{sec:discrete} and refer the reader to that section for further details, including notation, definitions of inner products, and discrete identities. We add, however, that each immersed point has an associated unit normal vector, comprising a vector of data $\normvec \in \vpoints$, and constructed to satisfy the discrete identity (\ref{eq:normident}).

Our ultimate objective is to develop discrete forms of the partial differential equations described in Section~\ref{sec:modelpdes}, including the associated single and double layers that constitute these equations augmented forms. We will start by developing discrete identities on differencing operators analogous to equations~(\ref{eq:gradf00}), (\ref{eq:lapphi}), and others in \ref{sec:genfun}. These identities will serve as the basis for much of the methodology that follows. First, however, we need to establish the discrete form of the masking functions, $\heavi(-\indic)$ and $\heavi(\indic)$.

\subsection{Discrete masking operators and calculus identities on the immersed points}
\begin{figure}[t]
\centering
\includegraphics[width=0.48\textwidth]{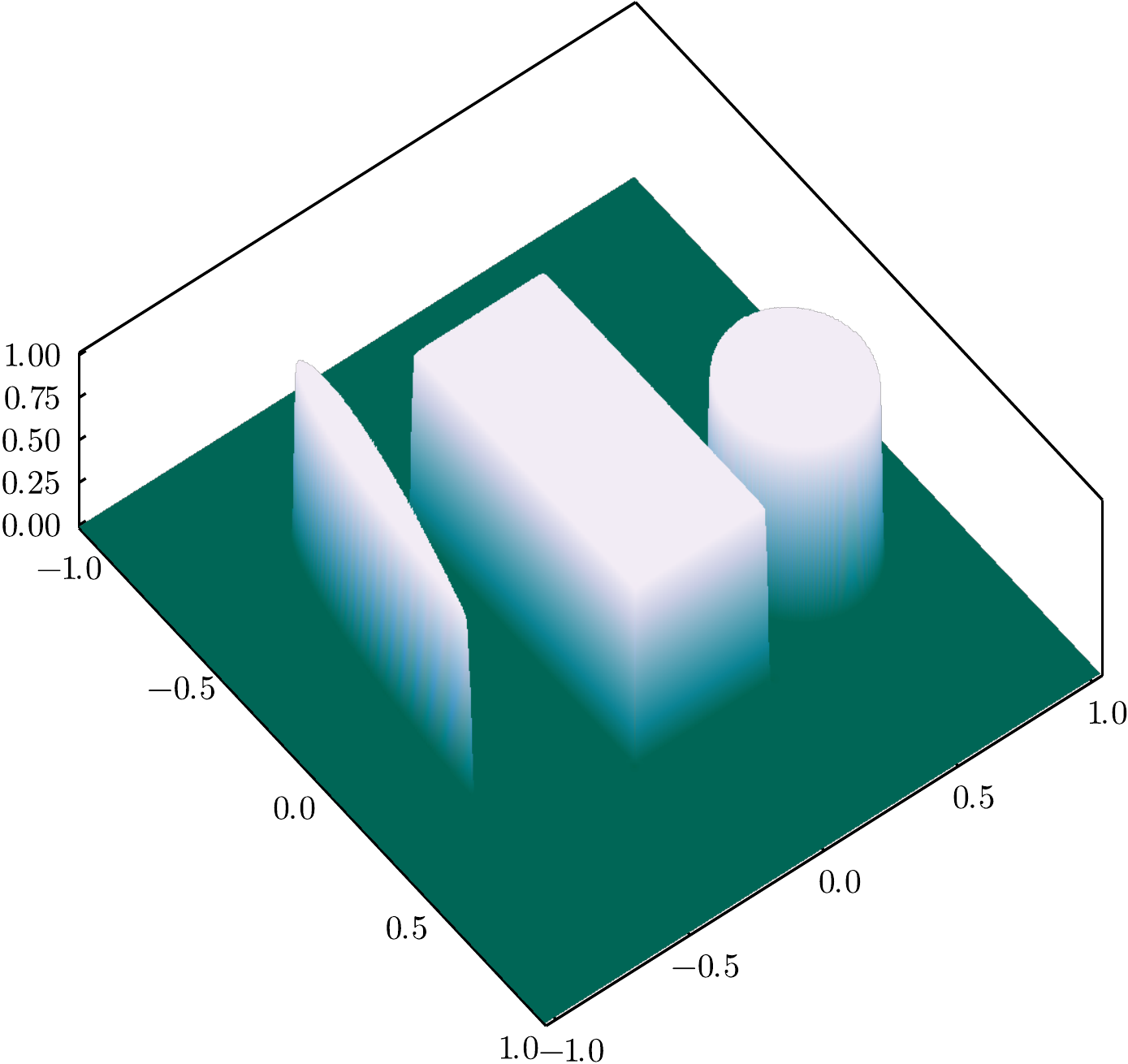}\includegraphics[width=0.48\textwidth]{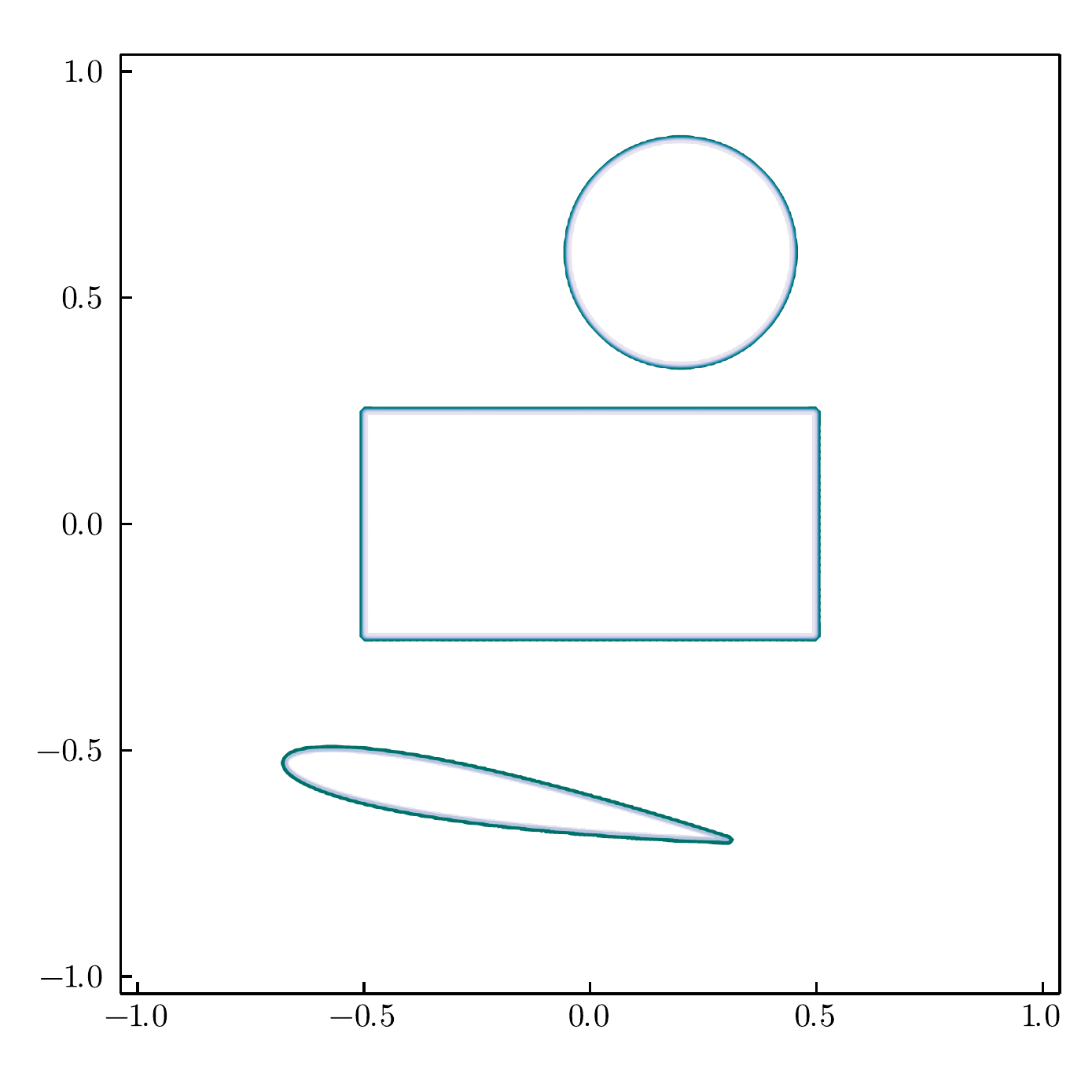}
\caption{Example of the interior masking operator $\hgridcenters^{-}$ for a set of two-dimensional shapes, using $\dx = 0.005$ and spacing between immersed points $\dS = 1.5\dx$.}\label{fig:masks}
\end{figure}

Our starting point for these will be to write the discrete analog of equation~(\ref{eq:gradheavi0}), but for the interior masking function $\heavi(-\indic)$, letting $\hgridcenters^{-}$ denote the discrete form of this function on the space of cell centers, $\centers$. The discrete equation we propose is
\begin{equation}
\label{eq:gradHgrid}
\gradgrid \hgridcenters^{-} = -\regds_{\faces} \normvec.
\end{equation}
The grouping on the right-hand side, $\regds_{\faces} \normvec$, which regularizes the surface normal vectors onto the cell faces on the grid, is a discrete version of $\dirac(\indic)\nrm$ and important in the rest of this paper. Equation (\ref{eq:gradHgrid}) does not quite give us a means of finding $\hgridcenters^{-}$, but if we take the discrete divergence of this equation, we get
\begin{equation}
\label{eq:poissonHgrid}
\lapgrid_{\centers} \hgridcenters^{-} = -\divgrid \regds_{\faces} \normvec,
\end{equation}
where $\divgrid$ is the discrete divergence. Thus, the masking operator is generated from source data comprising the divergence of the regularized normals of the discretized surface. The lattice Green's function on the cell centers and its associated inverse Laplacian $\invlapgrid_{\centers}$ allow us to immediately write the solution of (\ref{eq:poissonHgrid}): \begin{equation}
\label{eq:inheavigrid}
\hgridcenters^{-} = -\invlapgrid_{\centers} \divgrid \regds_{\faces} \normvec.
\end{equation}
Thus, we have formed the interior masking operator $\hgridcenters^{-}$ on the grid for a particular closed shape (or set of shapes), represented discretely by a set of immersed points and their associated areas and normals. An example of this masking operator, constructed for several shapes, is shown in Figure~\ref{fig:masks}. The exterior masking operator is easily obtained from the interior one, $\hgridcenters^{+} = \onesgrid - \hgridcenters^{-}$. The masking operators for vector-valued data, $\hgridfaces^{\pm} \in \faces$, are defined by interpolating the cell-centered functions:
\begin{equation}
\hgridfaces^{\pm} = \interpgrid{\centers}{\faces} \hgridcenters^{\pm}.
\end{equation}

The discrete Heaviside function defined in equation (\ref{eq:inheavigrid}) and the associated masking identities that will be developed in this section form the foundations of the present framework. It must be emphasized that, although we apply them using finite difference methods on a staggered Cartesian grid, these foundational equations are independent of the type of grid (structured or unstructured) or numerical method (finite difference, finite volume). Nevertheless, we now make use of some of the relationships available on a staggered grid to establish some helpful discrete identities for our current implementation. It can be readily shown that the divergence and Laplacian operations commute with one another on a staggered grid,
\begin{equation}
\label{eq:divlap}
\divgrid\lapgrid_{\faces} = \lapgrid_{\centers} \divgrid.
\end{equation}
If we right multiply by $\invlapgrid_{\faces}$ and left multiply by $\invlapgrid_{\centers}$, then this commutativity extends to the inverses, as well:
\begin{equation}
\divgrid\invlapgrid_{\faces} = \invlapgrid_{\centers} \divgrid.
\end{equation}
Thus, the interior masking function in (\ref{eq:inheavigrid}) can equivalently be written as
\begin{equation}
\label{eq:inheavigrid2}
\hgridcenters^{-} = -\divgrid \invlapgrid_{\faces}  \regds_{\faces} \normvec.
\end{equation}
This form is the discrete analog of (\ref{eq:interiormask}). Their difference in sign is due to the fact that the inverse of the discrete Laplacian is constituted by the negative of the lattice Green's function.

With the definitions presented thus far in this section, we can immediately obtain three crucial identities. The first arises when we apply the discrete gradient $\gradgrid$ to the masked vector
\begin{equation}
\label{eq:maskgridf}
\mask{\gridvar{\solnvec}}  = \gridvar{\solnvec}^{+}\had\hgridcenters^{+} + \gridvar{\solnvec}^{-}\had\hgridcenters^{-},
\end{equation}
where $\gridvar{\solnvec}^{\pm} \in \centers$. Using the first of a series of discrete product rules (\ref{eq:gradprod}) and the defining equation (\ref{eq:gradHgrid}) for the masking function, we get the discrete version of equation (\ref{eq:gradf00}) expressing the gradient of a masked scalar-valued grid vector:
\begin{equation}
\label{eq:gradgridmask}
\gradgrid \mask{\gridvar{\solnvec}} =  \mask{\gradgrid \gridvar{\solnvec}} + \interpgrid{\centers}{\faces} \left(\gridvar{\solnvec}^{+}-\gridvar{\solnvec}^{-}\right) \had \regds_{\faces}\normvec,
\end{equation}
where $\interpgrid{\centers}{\faces}$ is the interpolation operator from $\faces$ to $\centers$ and, for convenience, we have defined $\mask{\gradgrid \gridvar{\solnvec}}$ to denote the masked form of the gradient,
\begin{equation}
\mask{\gradgrid \gridvar{\solnvec}} \equiv  \left(\gradgrid \gridvar{\solnvec}^{+}\right) \had \hgridfaces^{+} + \left(\gradgrid \gridvar{\solnvec}^{-} \right) \had \hgridfaces^{-}.
\end{equation}

The second identity arises by applying the discrete divergence $\divgrid$ to the masked function
\begin{equation}
\mask{\gridvar{\vecvec}}  = \gridvar{\vecvec}^{+}\had\hgridfaces^{+} + \gridvar{\vecvec}^{-}\had\hgridfaces^{-},
\end{equation}
where $\gridvar{\vecvec}^{\pm} \in \faces$. Applying the second discrete product rule (\ref{eq:divprod}) and the definition (\ref{eq:gradHgrid}), we now get the discrete version of equation (\ref{eq:divf}) expressing the divergence of a masked vector-valued grid vector:
\begin{equation}
\divgrid \mask{\gridvar{\vecvec}} = \mask{\divgrid \gridvar{\vecvec}} + \interpgrid{\faces}{\centers}  \left(\left(\gridvar{\vecvec}^{+}-\gridvar{\vecvec}^{-}\right) \had \regds_{\faces}\normvec\right),
\end{equation}
where 
\begin{equation}
\label{eq:divgridmask}
\mask{\divgrid \gridvar{\vecvec}} \equiv  \left(\divgrid \gridvar{\vecvec}^{+}\right) \had \hgridcenters^{+} + \left(\divgrid \gridvar{\vecvec}^{-}\right) \had \hgridcenters^{-}.
\end{equation}
Though we do not present them, one can also generate similar identities for the discrete curl and gradient of vector-valued data and for the divergence of tensor-valued data, each of which also arises in the applications to follow.

Finally, by combining the identities (\ref{eq:gradgridmask}) and (\ref{eq:divgridmask}), it is straightforward to show that the discrete Laplacian applied to the masked vector (\ref{eq:maskgridf}) is
\begin{equation}
\lapgrid_{\centers} \mask{\gridvar{\solnvec}} = \mask{\lapgrid_{\centers}\gridvar{\solnvec}} + \interpgrid{\faces}{\centers} \left( \gradgrid(\gridvar{\solnvec}^{+}-\gridvar{\solnvec}^{-})\had\regds_{\faces}\normvec\right) + \divgrid\left( \interpgrid{\centers}{\faces} (\gridvar{\solnvec}^{+}-\gridvar{\solnvec}^{-}) \had \regds_{\faces}\normvec\right), \label{eq:extendedpoisson}
\end{equation}
where the masked Laplacian is defined as
\begin{equation}
\mask{\lapgrid_{\centers}\gridvar{\solnvec}}  = \left(\lapgrid_{\centers}\gridvar{\solnvec}^{+}\right) \had \hgridcenters^{+} + \left(\lapgrid_{\centers}\gridvar{\solnvec}^{-}\right) \had \hgridcenters^{-}.
\end{equation}
Equation (\ref{eq:extendedpoisson}) is the discrete equivalent of equation~(\ref{eq:lapphi}). 

\subsection{Immersed single and double layers}
\label{sec:immersedlayers}

Identities (\ref{eq:gradgridmask}), (\ref{eq:divgridmask}), and (\ref{eq:extendedpoisson}), analogous to their continuous counterparts, incorporate the jumps in field quantities across an interface into the discrete calculus operations on these field data. In particular, by simple comparison between (\ref{eq:extendedpoisson}) and the continuous equation (\ref{eq:lapphi}), the second term on the right-hand side of (\ref{eq:extendedpoisson}) resembles a discrete form of a single layer and the third term a discrete double layer. However, they use the grid itself to attach these jumps to the interface, so they are not in a form that readily allows us to introduce known data or prescribed conditions on the {\em interface} into the discrete operators. For such purposes, we would expect the discrete interface terms to each contain an element from the immersed point space $\spoints$ to describe the jump. Thus, we will define scalar-valued quantities on the sets of immersed points $\spoint{\slstrength}, \spoint{\dlstrength} \in \spoints$ as follows:
\begin{equation}
\label{eq:bcstrengths}
\regds_{\centers}\spoint{\slstrength} = \interpgrid{\faces}{\centers} \left(  \left(\gridvar{\vecvec}^{+}-\gridvar{\vecvec}^{-}\right) \had\regds_{\faces}\normvec\right),\qquad \regds_{\faces} \left( \spoint{\dlstrength} \had\normvec\right) = \interpgrid{\centers}{\faces} (\gridvar{\solnvec}^{+}-\gridvar{\solnvec}^{-}) \had \regds_{\faces}\normvec
\end{equation}
for $\gridvar{\vecvec}^{\pm} \in \faces$ and $\gridvar{\solnvec}^{\pm} \in \centers$. For brevity, we will refer to these terms as an {\em immersed single layer} and {\em immersed double layer}, respectively, and $\spoint{\slstrength}$ and $\spoint{\dlstrength}$ as their respective strengths.

To interpret these definitions, let us suppose that $\gridvar{\vecvec}^{\pm}$ and $\gridvar{\solnvec}^{\pm}$ respectively approximate vector-valued fields $\vecfunc^{\pm}$ and scalar-valued fields $\scafunc^{\pm}$,
\begin{equation}
\gridcomp{\vecvec^{\pm}}{\idex} \approx  \vecfunc^{\pm}(\xface{\idex}), \qquad\gridcomp{\solnvec^{\pm}}{\idex} \approx \scafunc^{\pm}(\xcenter{\idex}).
\end{equation}
We will denote the interface values of these fields by $\vecfunc_{b}^{\pm}$ and $\scafunc_{b}^{\pm}$. Definitions (\ref{eq:bcstrengths}) suggest that $\spoint{\slstrength}$ can be interpreted as the jump in the normal component of the interface values of $\vecfunc$, and that $\spoint{\dlstrength}$ can be interpreted as the jump in $\scafunc$ across the interface:
\begin{equation}
\spoint{\slstrength} = (\vecfunc_{b}^{+} - \vecfunc_{b}^{-})\cdot\nrm,\qquad \spoint{\dlstrength} = \scafunc_{b}^{+} - \scafunc_{b}^{-}.
\end{equation}
(Strictly, these equalities only hold at the discrete interface points, but we have omitted this for brevity.) In the generalized Laplacian in (\ref{eq:extendedpoisson}), the strength $\spoint{\slstrength}$ would represent the jump in the normal derivative of the scalar field.

Before we proceed to use these definitions to develop our solution procedures, it is useful to assess the error in the relationship between a grid-discretized field and its corresponding point-discretized interface values. One way to do this is to prescribe values for a scalar grid vector $\gridvar{\diffvec} \in \centers$, representing the difference $\gridvar{\solnvec}^{+} - \gridvar{\solnvec}^{-}$ on the grid, and to compare the results of two analogous operations on this vector: $\interpgrid{\centers}{\faces} \gridvar{\diffvec}\had \regds_{\faces}\normvec$, which is our immersed double layer defined above; and $\regds_{\faces}(\interp_{\centers} \gridvar{\diffvec}\had\normvec)$, which is the left-hand side of the double layer definition in (\ref{eq:bcstrengths}), but with the layer strength vector $\spoint{\dlstrength}$ replaced by the interpolation of $\gridvar{\diffvec}$ from the grid to the interface points. These two terms represent two ways of latching field data to the interface normals: one in which they are latched on the grid, and another in which they are first interpolated to the interface points, latched to the normals, and then regularized back to the grid.

\begin{figure}[t]
\centering
\includegraphics[width=0.7\textwidth]{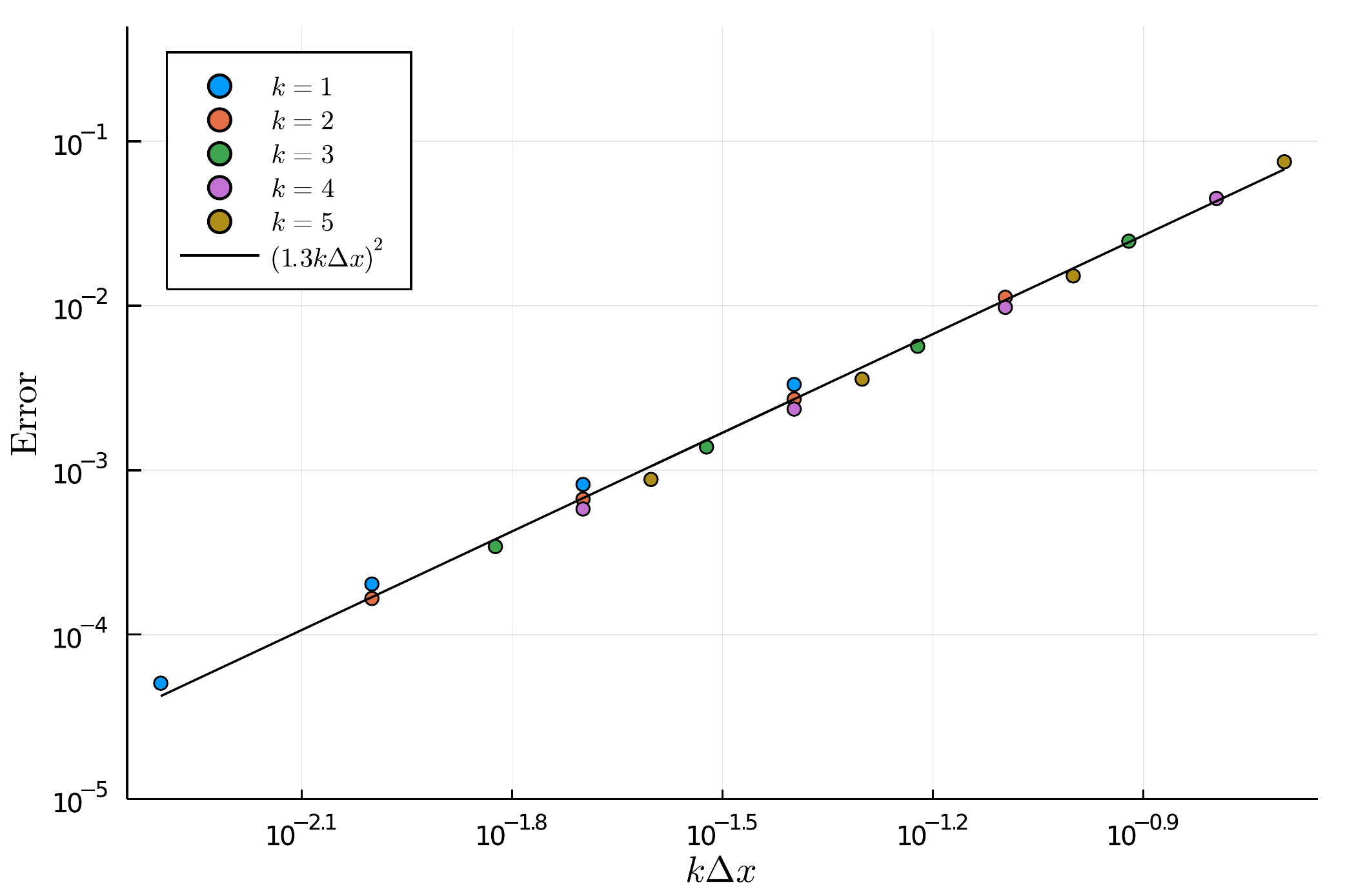}
\caption{Error $\normfaces{\interpgrid{\centers}{\faces} \gridvar{\diffvec}\had \regds_{\faces}\normvec - \regds_{\faces}(\interp_{\centers} \gridvar{\diffvec}\had\normvec)}/\normfaces{\interpgrid{\centers}{\faces} \gridvar{\diffvec}\had \regds_{\faces}\normvec}$ evaluated on a circle of radius 1, with $\gridvar{\diffvec} = \sin (\pi k x^{\centers})$, for various choices of $k$.}\label{fig:surfterm1} 
\end{figure}

For this comparison, we will evaluate these operations on the field $\gridvar{\diffvec} = \sin (\pi k x^{\centers})$, where $k$ and $\dx$ each take a variety of values, and the immersed points are arranged in a circular shape of radius $1$ and separated by distance $1.5\dx$ from each other. Throughout this paper, we use the smoothed 3-point DDF proposed by Yang et al.~\cite{yang2009smoothing}, though the results are similar for other DDF kernels, such as the one used by Roma et al.~\cite{roma99}. This comparison is depicted in Figure~\ref{fig:surfterm1}, in which the error (normalized by $\normfaces{\interpgrid{\centers}{\faces} \gridvar{\diffvec}\had \regds_{\faces}\normvec}$) is plotted versus the dimensionless grid spacing $k\Delta x$. The difference between these two analogous operations is $\sim (k\Delta x)^{2}$, determined by the order of accuracy of the grid interpolation $\interpgrid{\centers}{\faces}$. However, this does not imply that the error of the numerical solution of partial differential equations with the proposed method will be second-order accurate, but only that these two approaches of communicating data between the surface points and the grid are equivalent to within second-order accuracy. As we will see in the next section, the numerical solution of equations is first-order accurate.


\section{Numerical solution of prototype equations with immersed layers}
\label{sec:soln}

In this section we present the application of the immersed layers methodology to the three model problems discussed in Section~\ref{sec:modelpdes}: the two-dimensional Poisson equation and heat equation, each with Dirichlet boundary conditions, and the two-dimensional Navier--Stokes equations with moving rigid boundaries. All of the results described in this section have been obtained using the \verb+ViscousFlow+ package \cite{viscousflow}, available on GitHub. This package contains Jupyter notebooks that can be run to obtain solutions for each of these problems. 

\subsection{Solution of a Poisson problem with Dirichlet boundary conditions}

Suppose we seek to solve Poisson problems (\ref{eq:poissons}) in the regions exterior and interior to an interface defined by $\indic=0$, with Dirichlet boundary conditions $\varphi^{+} = \varphi^{+}_{b}$ and $\varphi^{-} = \varphi^{-}_{b}$ on the interface. In equation (\ref{eq:Dirichlet-matrix}) we described a means of setting up such a problem for solution in the continuous form. Here, we present a discrete analog of this problem and a means of solution. We approximate the continuous relationship between $\varphi$ and the forcing function $q$ in each region by discrete Poisson equations,
\begin{equation}
\lapgrid_{\centers}\gridvar{\solnvec}^{\pm} = \gridvar{\rhsvec}^{\pm}, 
\end{equation}
in which the scalar grid vector $\gridvar{\solnvec} \in \centers$ is the discrete approximation of $\varphi$ and $\gridvar{\rhsvec} \in \centers$ approximates $ q$:
\begin{equation}
\gridcomp{\solnvec}{\idex} \approx \varphi(\xcenter{\idex}),\qquad \gridcomp{\rhsvec}{\idex} \approx q(\xcenter{\idex}).
\end{equation}
 
 Let us write the masked form of grid vector $\gridvar{\solnvec}$,
\begin{equation}
\mask{\gridvar{\solnvec}} = \gridvar{\solnvec}^{+}\had\hgridcenters^{+} + \gridvar{\solnvec}^{-}\had\hgridcenters^{-},
\end{equation}
and an analogous one for $\gridvar{\rhsvec}$. Using (\ref{eq:extendedpoisson}) and the definitions of the immersed layers, the discrete Poisson equation for $\mask{\gridvar{\solnvec}}$ takes the form
\begin{equation}
\lapgrid_{\centers} \mask{\gridvar{\solnvec}} = \mask{\gridvar{\rhsvec}} +\regds_{\centers}\spoint{\slstrength}  + \divgrid \regds_{\faces} \left( \spoint{\dlstrength} \had\normvec\right), \label{eq:discretelayerpoisson}
\end{equation}
where the respective strengths are $\spoint{\slstrength} \in \spoints$ for the single layer and $\spoint{\dlstrength} \in \spoints$ for the double layer. Following our discussion in Section~\ref{sec:immersedlayers}, we should interpret $\spoint{\slstrength}$ as equal to the jump in the normal derivative of $\varphi$ across the interface and $\spoint{\dlstrength}$ as the jump in $\varphi$ itself. That is, the elements of $\spoint{\slstrength}$ approximate $(\grad\varphi_{b}^{+} - \grad\varphi_{b}^{-}) \cdot \nrm$, and those of $\spoint{\dlstrength}$ approximate $\varphi_{b}^{+}-\varphi_{b}^{-}$.


To solve for the masked vector $\mask{\gridvar{\solnvec}}$, we first set the double-layer strength $\spoint{\dlstrength}$ equal to the difference of the prescribed interface values,
\begin{equation}
\spoint{\dlstrength} = \spoint{\solnvec}^{+}_{b} - \spoint{\solnvec}^{-}_{b},
\end{equation}
where the components of these vectors are given by $\spoint{\solnvec}^{\pm}_{b}(\pindex) = \varphi^{+}_{b}(\X_{\pindex})$ for $\pindex=1,\ldots,N$. We also define $\overline{\spoint{\solnvec}}_{b} \in \spoints$ as the mean of these prescribed exterior and interior vectors,
\begin{equation}
\overline{\spoint{\solnvec}}_{b} = \frac{1}{2} \left( \spoint{\solnvec}^{+}_{b} + \spoint{\solnvec}^{-}_{b} \right),
\end{equation}
and, analogously to the continuous case, we constrain the interpolated solution vector on the immersed points to this mean:
\begin{equation}
\label{eq:mean-dirichlet}
\interp_{\centers} \mask{\gridvar{\solnvec}} = \overline{\spoint{\solnvec}}_{b}.
\end{equation}

Then, to solve the problem, we simultaneously seek the masked vector $\mask{\gridvar{\solnvec}}$ and the unknown single-layer strength vector $\spoint{\slstrength}$, which serves as a Lagrange multiplier for this constraint (\ref{eq:mean-dirichlet}). Our system of equations for these unknowns can be adapted from (\ref{eq:discretelayerpoisson}) and (\ref{eq:mean-dirichlet}) and written in matrix form as
\begin{equation}
\begin{bmatrix}
\lapgrid_{\centers} & \regds_{\centers} \\ \interp_{\centers} & 0
\end{bmatrix}
\begin{pmatrix}
\mask{\gridvar{\solnvec}} \\
-\spoint{\slstrength} 
\end{pmatrix} =
\begin{pmatrix}
\mask{\gridvar{\rhsvec}} + \divgrid \regds_{\faces} (\spoint{\dlstrength}\had\normvec) \\
\overline{\spoint{\solnvec}}_{b}
\end{pmatrix}. 
\end{equation}
We solve this saddle-point problem by means of a block-LU decomposition.

\begin{figure}[t]
\centering
\includegraphics[width=0.95\textwidth]{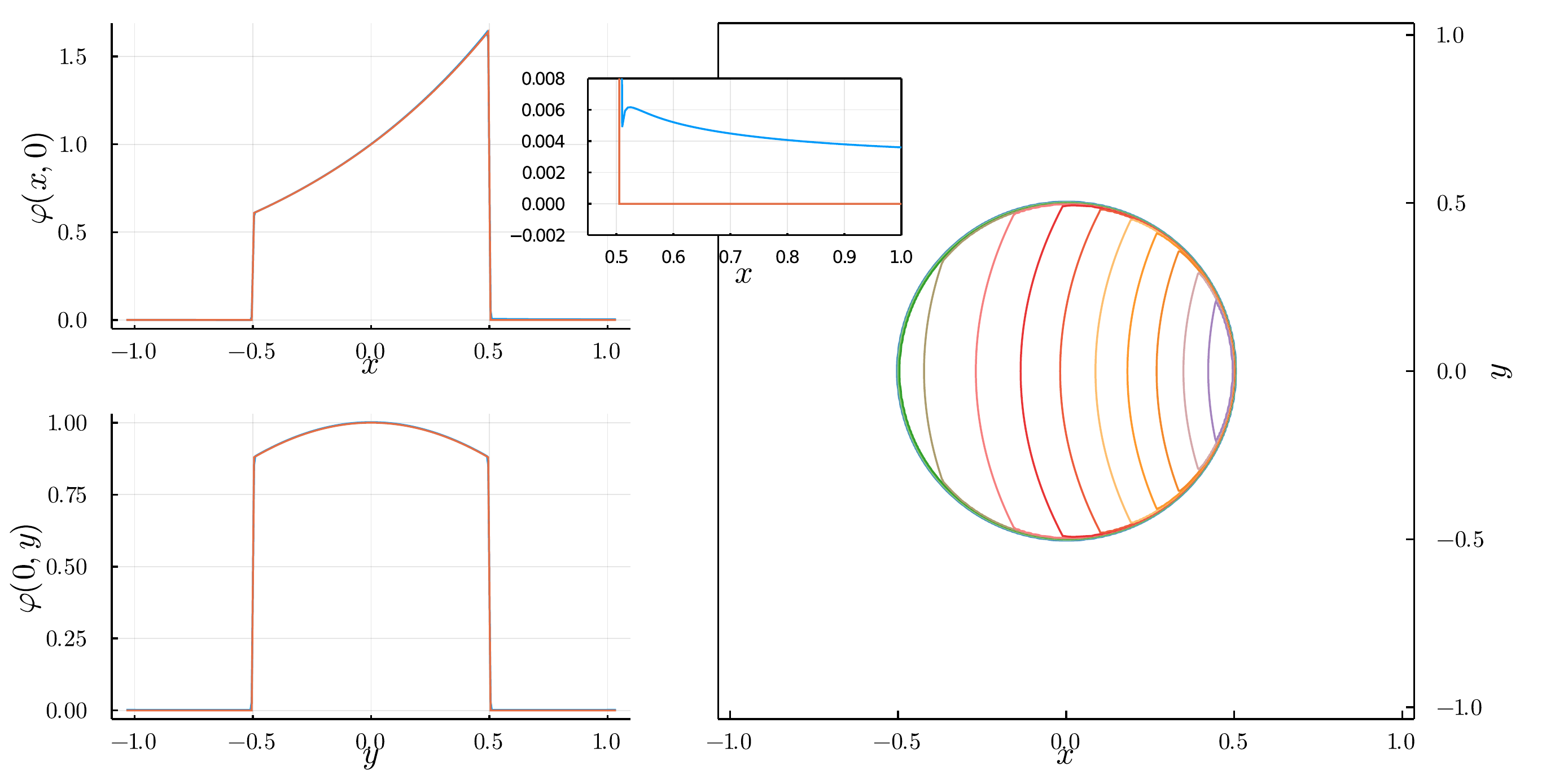}
\caption{(Left) Comparison between exact solution (red) and numerically-computed solution at $\dx/R= 0.01$ (blue) along $x$ and $y$ axes in example problem. (Right) Contour plot of numerical solution of Poisson example problem. (Inset) Detailed comparison of solutions just outside the interface.}
\label{fig:example1}
\end{figure}

\begin{figure}[tbp]
\centering
\includegraphics[width=0.48\textwidth]{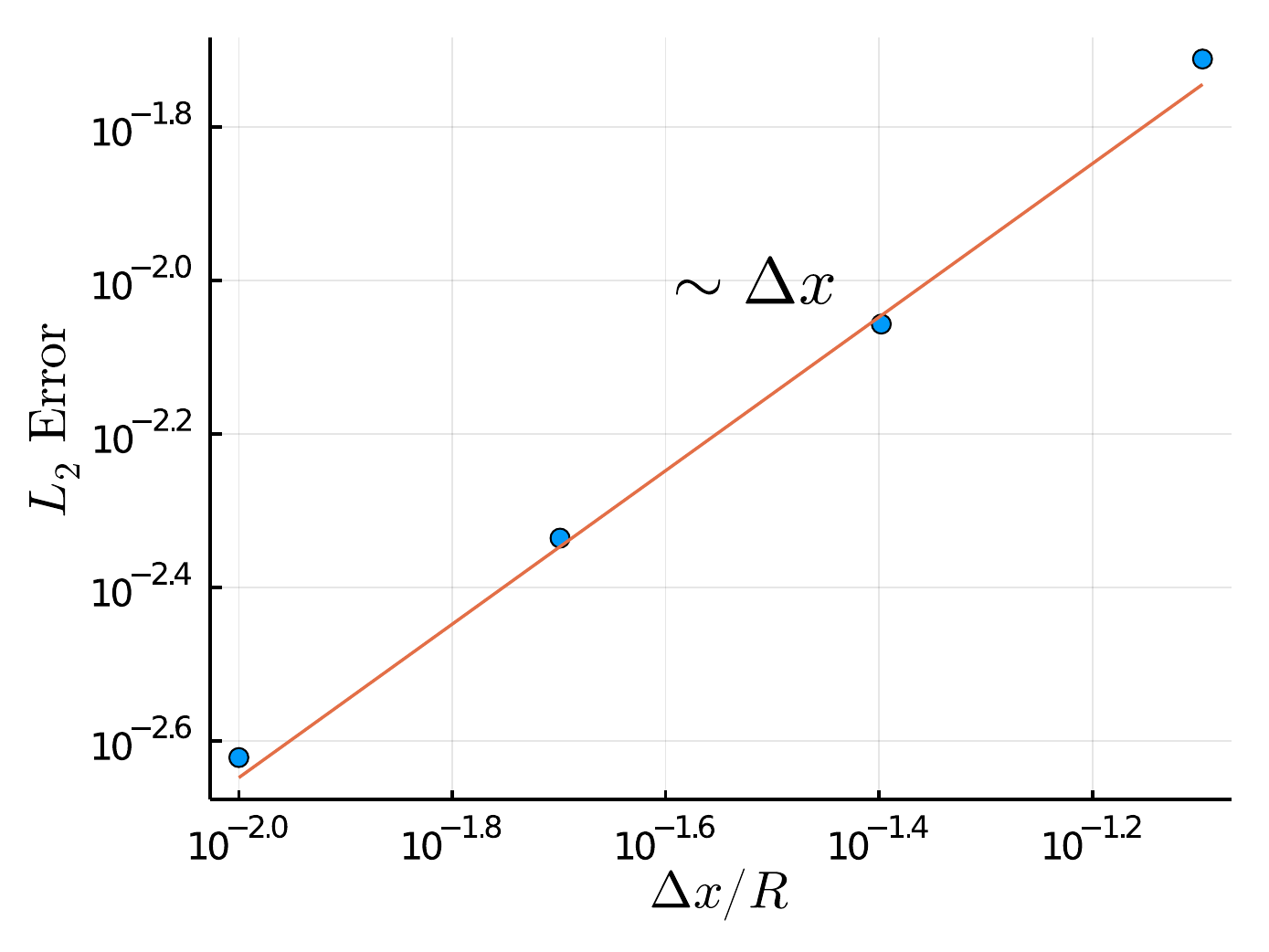}
\caption{$L_{2}$ norm of the error between exact solution and numerically-computed solution in Poisson example problem at various choices of grid spacing.}
\label{fig:example1err}
\end{figure}

As a demonstration of this methodology, we apply it to a two-dimensional example problem used by Leveque and Li \cite{levli94:1j}. The Laplace equation governs the respective solutions $\varphi^{+}$ and $\varphi^{-}$ in the regions exterior and interior to a circle of radius $R = 1/2$ centered at the origin. The conditions on this circle are different: $\varphi_{b}^{+} = 0$ and $\varphi_{b}^{-} = \mathrm{e}^{x}\cos y$. It is easy to verify that these are also the solutions throughout their respective regions. In Figure~\ref{fig:example1} we show the solution of this problem with $\dx/R = 0.01$ and with the circular interface discretized with points separated by a distance of approximately $1.5\dx$ (that is, $N = 418$). It is clear that the solution matches well in both the interior and exterior regions with the exact solution. In particular, the numerical solution in the superfluous region, exterior to a zone surrounding the interface of approximately two cells' width, is smaller than 1 percent. In Figure~\ref{fig:example1err} we show that this error (measured here by the $L_{2}$ norm) converges to zero at the expected rate, proportional to $\Delta x$. 

\subsection{Convection-diffusion equation with Dirichlet boundary conditions}

In this section, we apply our discrete operators to the solution of the convection-diffusion equation (\ref{eq:convdiff}). This discrete form, for a scalar field $\gridvar{\solnvec}\in \centers$ convected by a velocity field $\gridvar{\velvec} \in \faces$, is obtained in much the same the same way that we used to develop the Poisson equation: using the extended form of discrete Laplace operator (\ref{eq:extendedpoisson}) and then replacing the surface terms with the immersed single and double layers defined in (\ref{eq:bcstrengths}). Additionally, we discretize the convective term using the form that satisfies the discrete product rule (\ref{eq:divprod}). We arrive at a form that mirrors the continuous form (\ref{eq:maskedconvdiff}),
\begin{equation}
\ddt{\mask{\gridvar{\solnvec}}} + \interpgrid{\faces}{\centers}\left(  \mask{\gridvar{\velvec}} \had \gradgrid \mask{\gridvar{\solnvec}}\right) = \kappa \lapgrid_{\centers} \mask{\gridvar{\solnvec}} + \mask{\gridvar{q}} + \regds_{\centers} \spoint{\slstrength} - \divgrid\regds_{\faces} \left( \kappa \spoint{\dlstrength}\had\normvec\right),
\end{equation}
where $\mask{\gridvar{q}} \in \centers$ is a known forcing vector and
\begin{equation}
\spoint{\dlstrength} = \spoint{\solnvec}^{+}_{b} - \spoint{\solnvec}^{-}_{b}.
\end{equation}
As in the case of the Poisson equation, the single-layer strength vector $\spoint{\slstrength} \in \spoints$ serves as a Lagrange multiplier for the Dirichlet condition,
\begin{equation}
\interp_{\centers}\mask{\gridvar{\solnvec}} = \mask{\spoint{\solnvec}}_{b},
\end{equation}
where
\begin{equation}
\mask{\spoint{\solnvec}}_{b} = \frac{1}{2} \left( \spoint{\solnvec}^{+}_{b} + \spoint{\solnvec}^{-}_{b} \right).
\end{equation}
By equation (\ref{eq:sforcedef}), the constraint force $\spoint{\slstrength}$ can be interpreted as an approximation of the local diffusive flux of $\varphi$ into the circle. 

As a demonstration, we evaluate the methodology for solving the two-dimensional homogeneous linear diffusion equation (with convective velocity set to zero) in a circle of radius $R$, with Dirichlet boundary conditions on the circle. For the interior of the circle, we set the Dirichlet value on the circle to a uniform constant, $\varphi^{-}_{b} = 1$; exterior to the circle, the Dirichlet value is set to zero, $\varphi^{+}_{b} = 0$. The exact solution of this problem in the circle interior is \citep{carslaw1959conduction}
\begin{equation}
\varphi^{-}(r,t) = 1 - \sum_{k=1}^{\infty} \frac{2}{j_{k} J_{1}(j_{k})}  \mathrm{e}^{-j_{k}^{2} \kappa t/R^{2} } J_{0}(j_{k}r/R),
\end{equation}
where $r = (x^{2}+y^{2})^{1/2}$, $\kappa$ is the diffusion coefficient, $J_{n}$ is the $n$th-order Bessel function of the first kind, and $j_{k}$ is the $k$th root of $J_{0}(j_{k}) = 0$. The exact solution outside the circle remains zero, $\varphi^{+} = 0$, for all time.

For the numerical solution, we set $\spoint{\solnvec}^{+}_{b} = 0$ and $\spoint{\solnvec}^{-}_{b} = \onesspoint$. The grid solution vector $\mask{\gridvar{\solnvec}}$ is initially set to zero; grid velocity $\mask{\gridvar{\velvec}}$ is absent in this problem. The equations can be written as
\begin{equation}
\begin{bmatrix}\displaystyle
\lindiff {\centers}{\kappa}   & \regds_{\centers}  \\ \displaystyle
\interp_{\centers} & 0 
\end{bmatrix}
\begin{pmatrix}
\mask{\gridvar{\solnvec}} \\  -\spoint{\slstrength}
\end{pmatrix} = 
\begin{pmatrix} 
\mask{\gridvar{q}}  - \divgrid\regds_{\faces} \left( \kappa \spoint{\dlstrength}\had\normvec\right)   \\ \mask{\spoint{\solnvec}}_{b}
\end{pmatrix}.
\end{equation}
where $\mask{\spoint{\solnvec}}_{b} = \frac{1}{2}\onesspoint$, $\spoint{\dlstrength} = -\onesspoint$, and we have defined $\lindiff {\centers}{\kappa}$, the semi-discrete linear diffusion operator acting at cell centers, as
\begin{equation}
\label{eq:lindiff}
\lindiff {\centers}{\kappa} \equiv \ddt {} -  \kappa \lapgrid_{\centers}.
\end{equation}
Writing the equations in this form reveals the underlying saddle-point structure of the problem and also renders it conducive to solution by the 2nd-order integrating factor/half-explicit Runge--Kutta method developed by Liska and Colonius \cite{liska16}. At the end of the $n$th time step, the solution, $\mask{\gridvar{\solnvec}}^{n}$, and the associated constraint force, $\spoint{\slstrength}^{n}$, are solved for simultaneously. In each case considered below, the time step is chosen so that $\kappa\Delta t/\dx^{2} = 0.5$.

\begin{figure}[t]
\centering
\includegraphics[width=\textwidth]{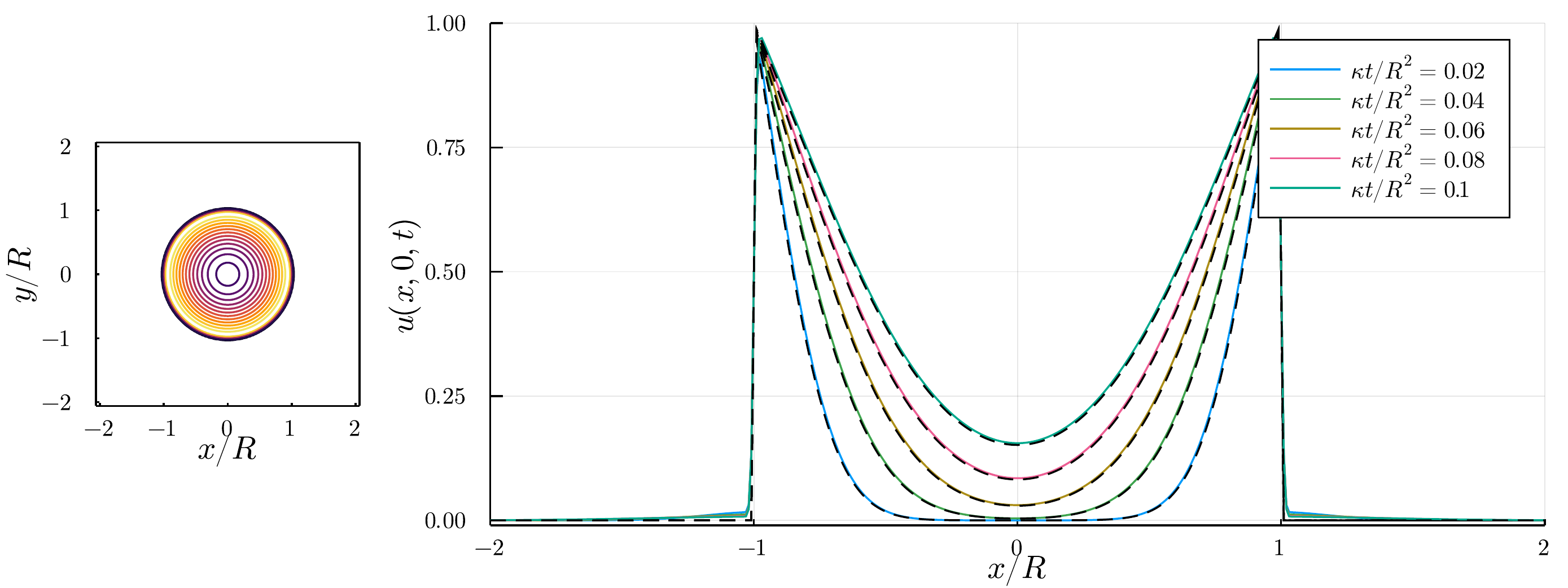}
\caption{Diffusion example problem.  (Left) Numerical solution of at $\kappa t/R^{2} = 0.1$, using $\Delta x/R = 0.01$. (Right) Comparison between numerical solution with $\Delta x/R = 0.01$ and exact solution (black dashed lines) at several instants along the line $y=0$.}
\label{fig:example2}
\end{figure}

\begin{figure}[t]
\centering
\includegraphics[width=0.48\textwidth]{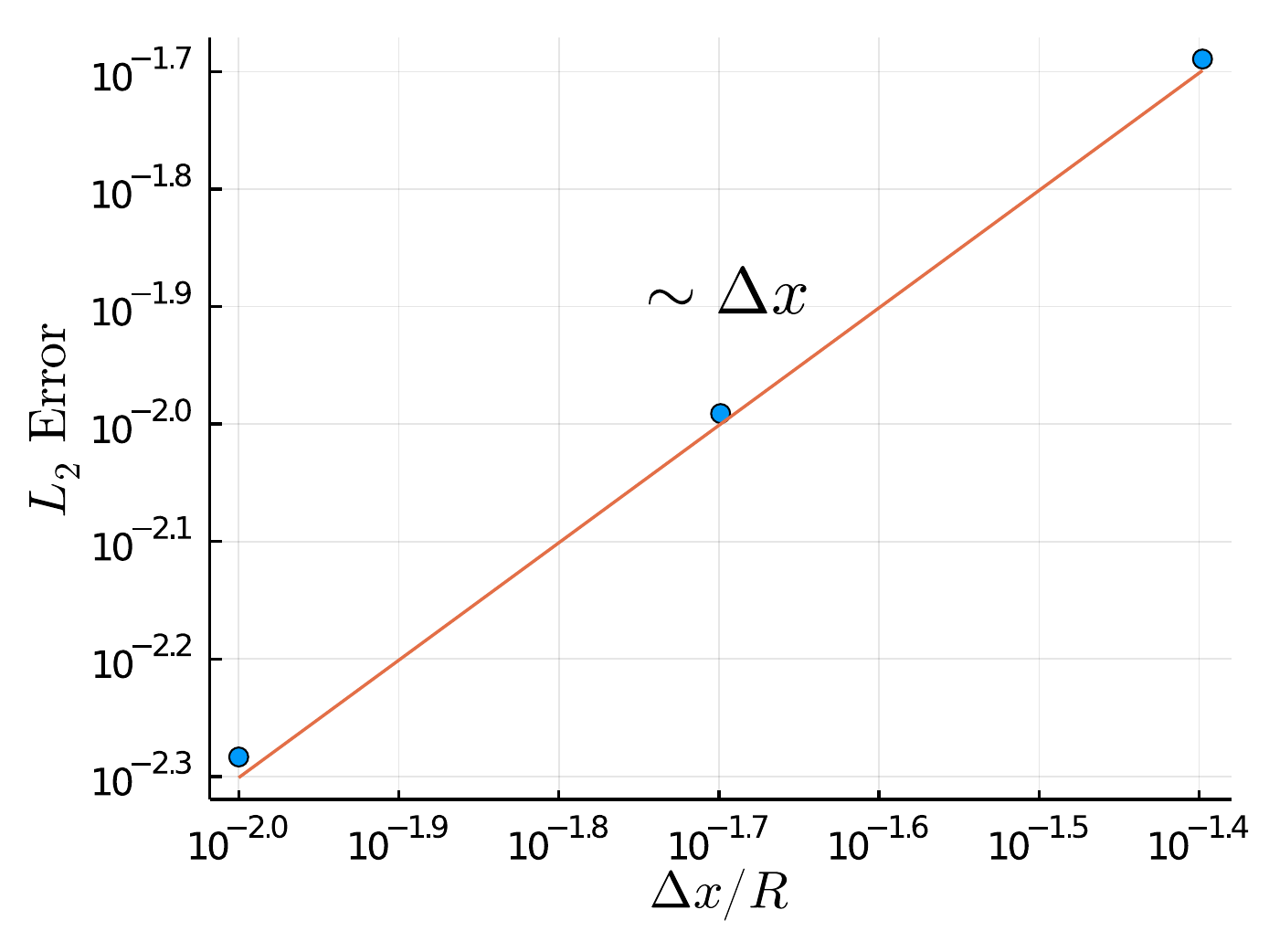} 
\caption{For diffusion problem, $L_{2}$ norm of the error between the numerical solution and exact solution at $\kappa t/R^{2} = 0.1$.}
\label{fig:example2err}
\end{figure}

The solution of the problem for grid spacing $\dx/R = 0.01$ is shown in Figure~\ref{fig:example2} and compared with the exact solution at several instants. In these results, we can see that the numerical solution is nearly zero outside of the circle except for a small region adjacent to the set of immersed points. Importantly, this region of non-zero values does not vary substantially with time, i.e., any spurious diffusion outside the circle is insignificant. When we calculate the normalized $L_{2}$ error of the solution at a single time instant with a variety of grid spacings, shown in the left panel of Figure~\ref{fig:example2err}, we find first-order accuracy as the grid spacing decreases.


\subsection{Application to the Navier--Stokes equations}

In this section we apply the immersed layer approach to approximate the Navier--Stokes equations (\ref{eq:maskedns}). Denoting the grid approximation of the masked velocity field by $\mask{\gridvar{\velvec}}$, we can write the semi-discrete Navier--Stokes equations as
\begin{subequations}
\begin{align}
\ddt{\mask{\gridvar{\velvec}}} + N(\mask{\gridvar{\velvec}}) &= -\gradgrid \mask{\gridvar{\press}} + \nu \lapgrid_{\faces} \mask{\gridvar{\velvec}} - \regds_{\faces} \vpoint{\vslstrength} - \divgrid_{\tensors} \regds_{\tensors} \tpoint{S},  \label{eq:gridns}\\
\divgrid \mask{\gridvar{\velvec}}&= \regds_{\centers} \left(\vpoint{\vdlstrength}\cdot\normvec\right), \label{eq:gridcont}\\
\interp_{\faces} \mask{\gridvar{\velvec}} &=\mask{\vpoint{\velvec}}_{b}, \label{eq:gridnoslip}
\end{align}
\end{subequations}
where $\mask{\gridvar{\press}}$ is a grid pressure, whose role is to enforce the discrete continuity equation (equation (\ref{eq:gridcont})); $N(\mask{\gridvar{\velvec}})$ is an approximation of the convective acceleration, using the first term on the right-hand side of (\ref{eq:discretedivuv}); and we have defined
\begin{equation}
\vpoint{\vdlstrength} =\vpoint{\velvec}^{+}_{b} - \vpoint{\velvec}^{-}_{b}, \qquad \mask{\vpoint{\velvec}}_{b} =\frac{1}{2}\left(\vpoint{\velvec}^{+}_{b} + \vpoint{\velvec}^{-}_{b}\right), \qquad \tpoint{S} = \nu \left( \vpoint{\vdlstrength}\tensprod \normvec + \normvec \tensprod\vpoint{\vdlstrength} \right),
\end{equation}
The surface tensor $\tpoint{S}$ is the discrete form of the viscous surface tensor $\boldsymbol{\Sigma}$, defined in (\ref{eq:viscoussurf}). We can write these equations in saddle-point form as
\begin{equation}
\begin{bmatrix}
\lindiff{\faces}{\nu} & \gradgrid & \regds_{\faces} \\
\divgrid & 0 & 0 \\
\interp_{\faces} & 0 & 0
\end{bmatrix}
\begin{pmatrix}
\mask{\gridvar{\velvec}} \\ \mask{\gridvar{\press}} \\ \vpoint{\vslstrength}
\end{pmatrix} = 
\begin{pmatrix}
-N(\mask{\gridvar{\velvec}})  - \divgrid_{\tensors} \regds_{\tensors} \tpoint{S}   \\ \regds_{\centers} \left(\vpoint{\vdlstrength}\cdot\normvec\right) \\ \mask{\vpoint{\velvec}}_{b}
\end{pmatrix},
\end{equation}
using a definition for the linear diffusion operator $\lindiff{\faces}{\nu}$ similar to that in (\ref{eq:lindiff}).

Extending our interpretation from the continuous form (\ref{eq:surfaceflux}), the Lagrange multiplier vector $\vpoint{\vslstrength} \in \vpoints$ is a discrete approximation of the jumps in surface traction and momentum flux  (scaled by density).  The force and moment on a body are then straightforward to compute. For example, on a single body with external surface velocity $\vpoint{\velvec}^{+}_{b} = \vpoint{\velvec}_{b}$ and with internal velocity set to zero to ensure zero spurious motion and stress inside the body, the traction exerted by the fluid follows from the analogy with (\ref{eq:tractionext})
\begin{equation}
\vpoint{\vslstrength} - \frac{1}{2}\vpoint{\velvec}_{b}\tensprod \vpoint{\velvec}_{b}\cdot \normvec
\end{equation}
and the $k$th component of the density-scaled force is 
\begin{equation}
f_{k} = \ipvectorbig{\onesspoint^{(k)}}{\vpoint{\vslstrength} - \frac{1}{2}\vpoint{\velvec}_{b}\tensprod \vpoint{\velvec}_{b}\cdot \normvec}.  
\end{equation}
The moment $m_{k}$ is straightforward, utilizing a moment arm generated by the vector of point coordinates $\xpointvec \in \vpoints$,
\begin{equation}
m_{k} = \ipvectorbig{\onesspoint^{(k)}}{\xpointvec \times \left(\vpoint{\vslstrength} - \frac{1}{2}\vpoint{\velvec}_{b}\tensprod \vpoint{\velvec}_{b}\cdot \normvec\right)}.  
\end{equation}

If we take the discrete curl of the masked velocity, we get the equivalent of (\ref{eq:vorttot}), a discrete vorticity $\gridvorttot \in \edges$ that includes both the masked vorticity as well as a discrete jump across the interface points:
\begin{equation}
\gridvorttot = \rotgrid \mask{\gridvar{\velvec}} = \mask{\gridvar{\vortvec}} + \regds_{\edges} (\normvec \times \vpoint{\vdlstrength}).
\end{equation}
By taking the curl of (\ref{eq:gridns}), and exploiting the commutativity of the discrete operators, we arrive at the discrete vorticity transport equation,
\begin{subequations}
\begin{align}
\ddt{\gridvorttot} + \rotgrid\regds_{\faces} \vpoint{\vslstrength}  &= -\rotgrid N(\mask{\gridvar{\velvec}}) + \nu \lapgrid_{\edges} \gridvorttot  - \rotgrid\divgrid_{\tensors} \regds_{\tensors} \tpoint{S}  \\
\interp_{\faces} \mask{\gridvar{\velvec}} &=\mask{\vpoint{\velvec}}_{b}
\end{align}
\end{subequations}

We reconstruct the masked velocity field from the discrete equivalent of the Helmholtz decomposition (\ref{eq:helm})
\begin{equation}
\mask{\gridvar{\velvec}} = \curlgrid \mask{\gridvar{\sfvec}} + \gradgrid \mask{\gridvar{\potvec}} + \gridvar{U}_{\infty},
\end{equation}
in which the discrete potentials $\mask{\gridvar{\sfvec}} \in \edges$ and $\mask{\gridvar{\potvec}} \in \centers$ each are the solution of a Poisson equation,
\begin{equation}
\lapgrid_{\edges} \mask{\gridvar{\sfvec}} = -\gridvorttot,\qquad \lapgrid_{\centers} \mask{\gridvar{\potvec}} = \regds_{\centers} \left(\vpoint{\vdlstrength}\cdot\normvec\right).
\end{equation}
Thus, we can write the overall system of equations as
\begin{equation}
\begin{bmatrix}
\lindiff{\edges}{\nu}  & \rotgrid\regds_{\faces} \\
-\interp_{\faces} \curlgrid\invlapgrid_{\edges} & 0 &
\end{bmatrix}
\begin{pmatrix}
\gridvorttot \\  \vpoint{\vslstrength}
\end{pmatrix} = 
\begin{pmatrix}
-\rotgrid N(\mask{\gridvar{\velvec}})  - \rotgrid\divgrid_{\tensors} \regds_{\tensors} \tpoint{S}  \\ \mask{\vpoint{\velvec}}_{b} - \interp_{\faces} \gradgrid \invlapgrid_{\centers} \regds_{\centers} \left(\vpoint{\vdlstrength}\cdot\normvec\right) - \interp_{\faces} \gridvar{U}_{\infty}
\end{pmatrix},
\end{equation}
where $\lindiff{\edges}{\nu}$ is defined similarly to (\ref{eq:lindiff}) but with viscosity $\nu$ and on the space of cell edges, $\edges$.

\begin{figure}[t]
\centering
\includegraphics[width=0.48\textwidth]{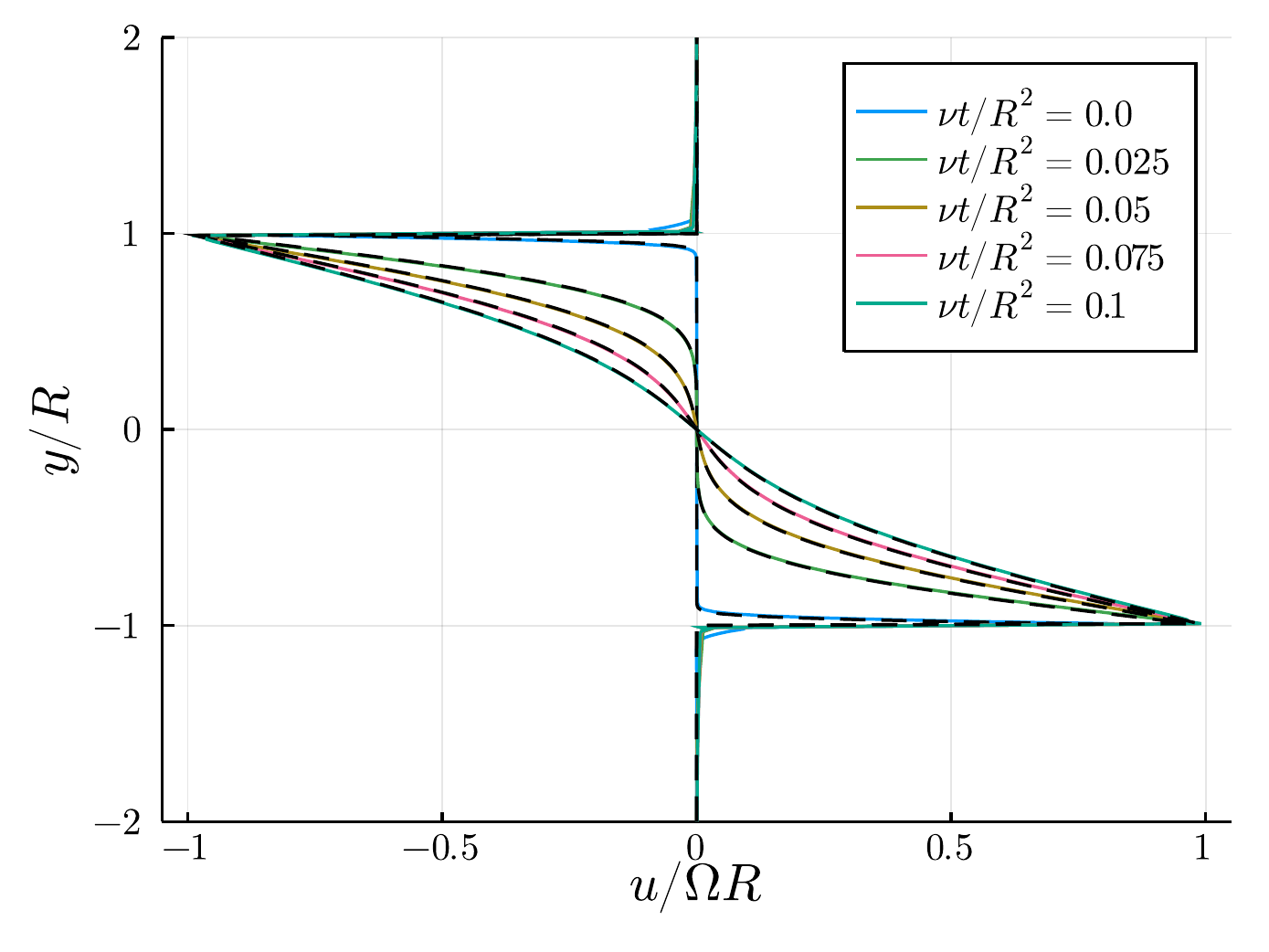}\quad \includegraphics[width=0.48\textwidth]{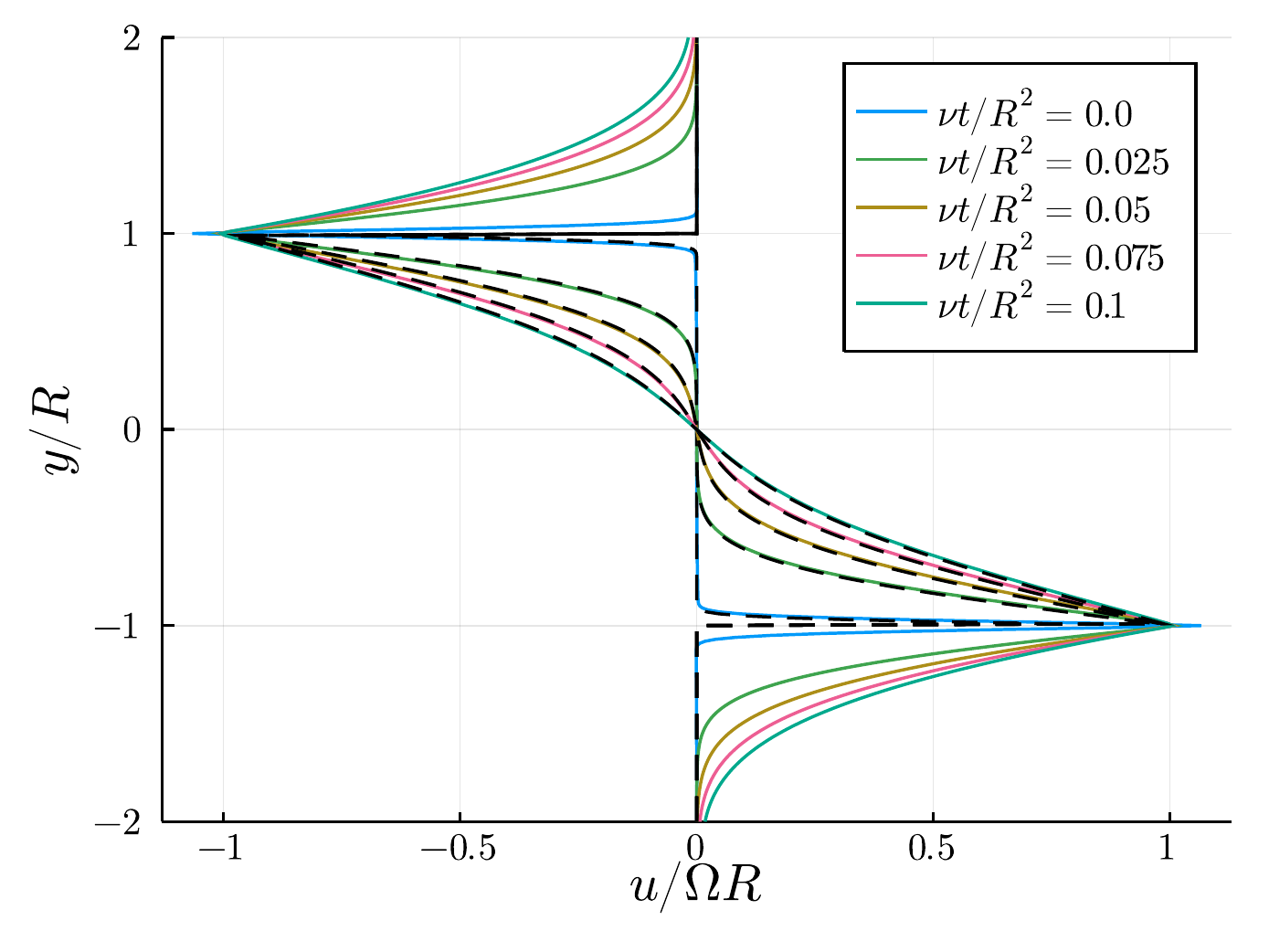}
\caption{Profiles of $x$ component of velocity along line $x=0$ inside impulsively-rotated circular region at Reynolds number 100 at various instants. Exact solution shown as dashed line at each instant. Left: numerical solution carried out with the present method with $\dx/R = 0.01$. Right: solution with IBPM \cite{liskacolonius17} with $\dx/R = 0.01$.}\label{fig:rotatecircvel}
\end{figure}

\begin{figure}[t]
\centering
\includegraphics[width=0.48\textwidth]{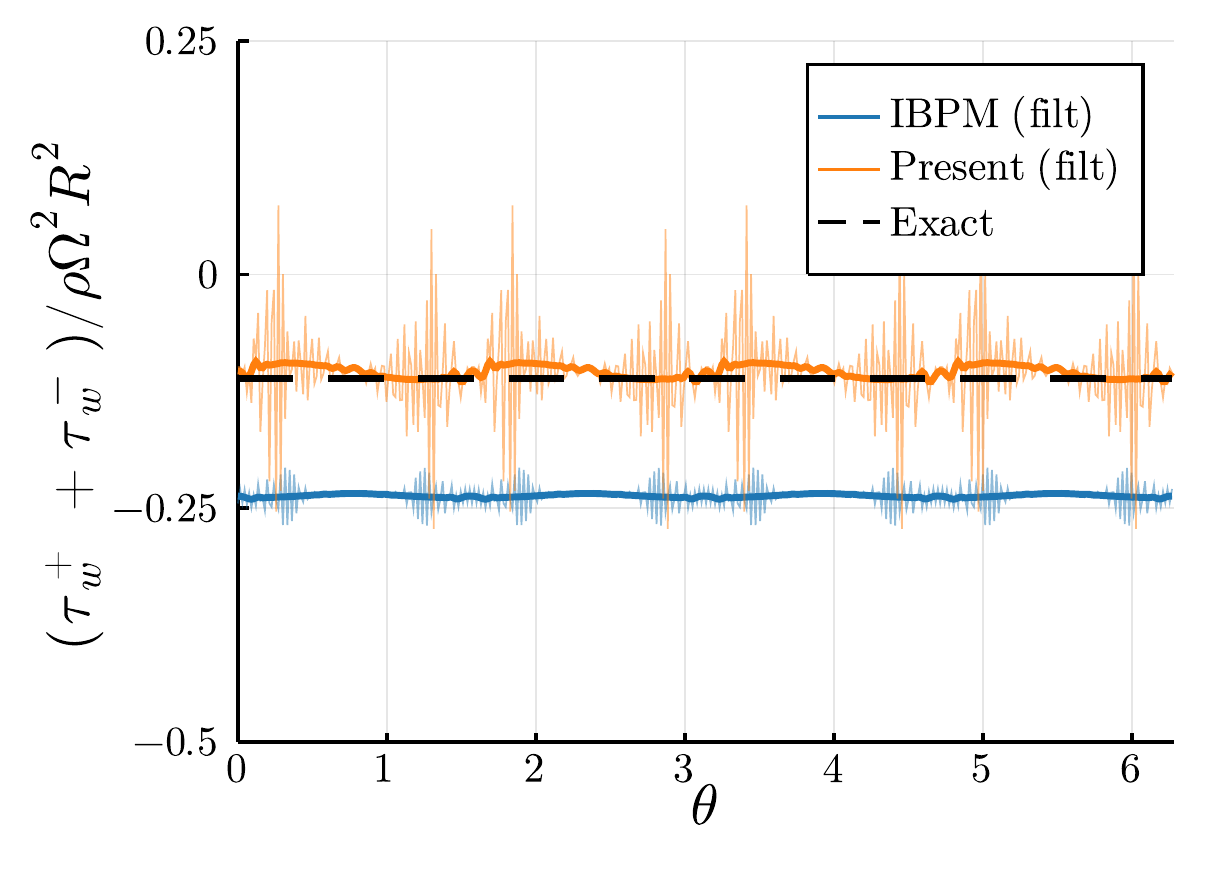}
\includegraphics[width=0.48\textwidth]{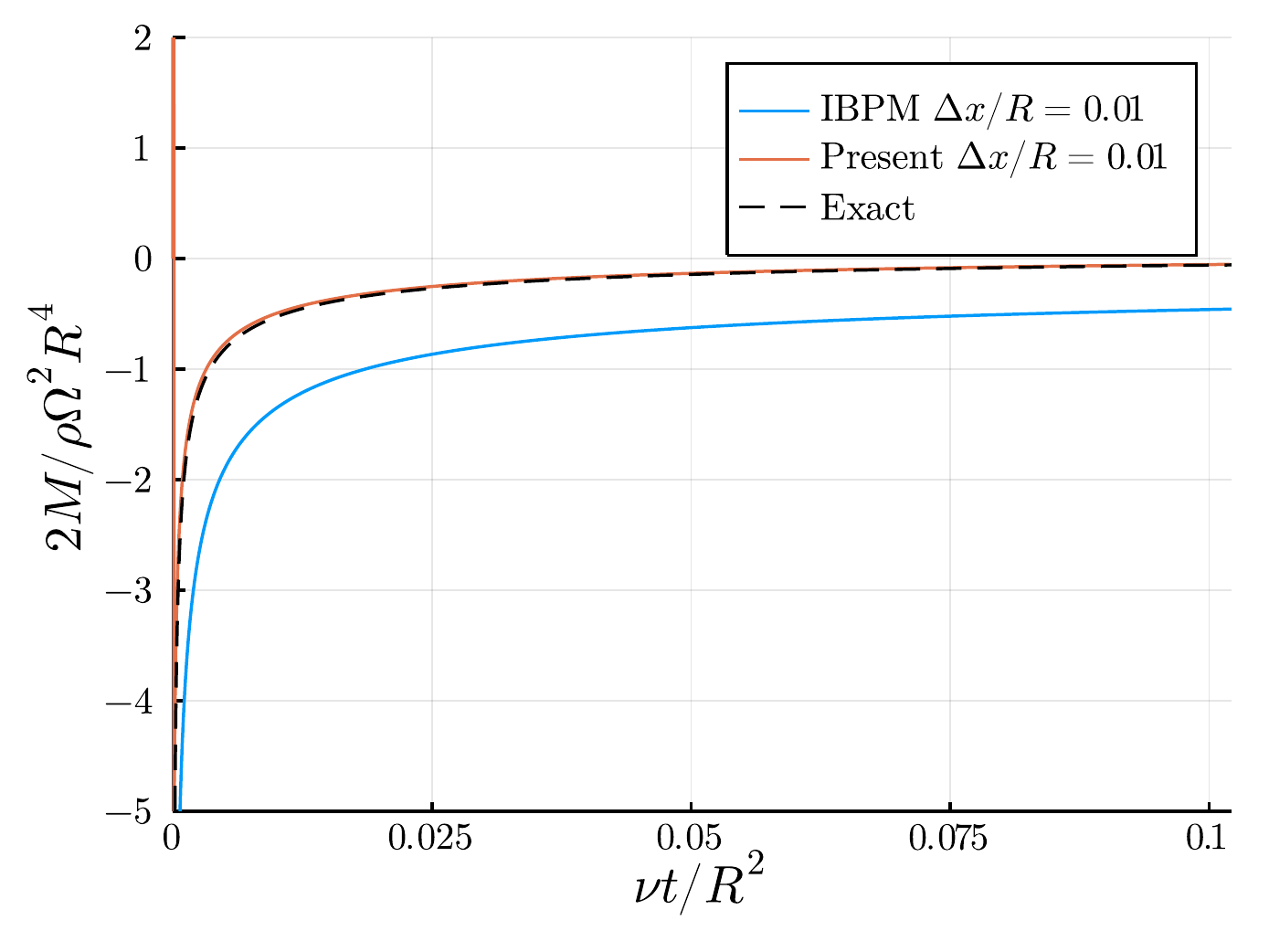}
\caption{Comparison of wall shear stress at $\nu t/R^{2} = 0.002$ (right) and moment (left) applied by fluid on the wall from impulsive rotation of a circular region of fluid. The shear stress distributions depict both the original predictions (in light color) as well as the distributions after applying the filtering technique of Goza et al.~\cite{goza2016accurate} (darker lines).}\label{fig:rotatecircmom}
\end{figure}

As in the diffusion problem, we utilize a 2nd-order half-explicit Runge--Kutta method with integrating factor \cite{liska16} to advance the system of equations. To validate the methodology, we present the application on a problem with an exact solution and with non-zero wall velocity: the two-dimensional fluid motion inside a circular region rotated impulsively about its center. We denote the constant angular velocity by $\Omega$ and the radius of the region by $R$. It is straightforward to show that the exact solution of this flow is given by the azimuthal velocity and moment
\begin{equation}
v_{\theta}(r,t) = \Omega R \left( \frac{r}{R} - 2 \sum_{n=1}^{\infty} \frac{J_{1}(\lambda_{n}r/R) }{J_{2}(\lambda_{n})} \mathrm{e}^{-\lambda_{n}^{2}\nu t/R^{2}}\right), \qquad M_{z} = 4\pi \Omega R^{2} \sum_{n=1}^{\infty} \mathrm{e}^{-\lambda_{n}^{2}\nu t/R^{2}},
\end{equation} 
respectively, where $\lambda_{n}$ are the zeros of the first-order Bessel function of the first kind.

Results for the numerical solution of this problem using the present method at Reynolds number $\Omega R^{2}/\nu = 100$ are compared with this exact solution in Figure~\ref{fig:rotatecircvel}. Goza et al.~\cite{goza2016accurate} simulated this problem with the IBPM, necessarily generating flows both internal and external to the rotating circle; for reference we show the results obtained with the IBPM, using the lattice Green's function version of this method developed by Liska and Colonius \cite{liskacolonius17}. In both numerical solutions, the grid spacing is $\dx/R = 0.01$, the time step size is set so that $\nu\Delta t/\Delta x^{2} = 0.5$ (so that $\Omega \Delta t = 0.005$), and the points on the circular wall are spaced by $1.5\dx$.  The agreement between both numerical methods and the exact solution is very good in the interior of the circle. However, as expected, the IBPM also generates a superfluous flow outside the circle. The effect of this superfluous flow is exhibited in Figure~\ref{fig:rotatecircmom}. The left panel of this figure depicts the sum of the internal and external wall shear stress distributions (i.e., the tangential component of $\vpoint{\vslstrength}$) obtained by each method at $\nu t/R^{2} = 0.002$ ($\Omega t = 0.2$). First, we note that the distribution predicted by the present method is somewhat noisier than that predicted by the IBPM. This noise, explored by Goza et al.~\cite{goza2016accurate}, is due to the ill conditioning of the Schur complement system, which tends to amplify high-frequency noise contained in the right-hand side vector of this system. Since the Schur complement operators are the same in the two methods, the origin of the larger noise in the present method lies in the right-hand side (in the terms associated with the jumps in fluid velocity). However, after applying the filtering technique devised in \cite{goza2016accurate} (which we emphasize is an optional post-processing step) we obtain a cleaner distribution, as shown in the figure. In contrast to the IBPM, the present method only contains the influence of the internal flow, so its value agrees well with the exact solution, whereas the IBPM exhibits the sum of shear stresses from both the internal and external flows. This agreement is revealed more clearly by the methods' moment histories, shown in the right panel, where the present method's prediction remains close to the exact solution for the entire interval.

\begin{figure}[tb]
\centering
\includegraphics[width=\textwidth]{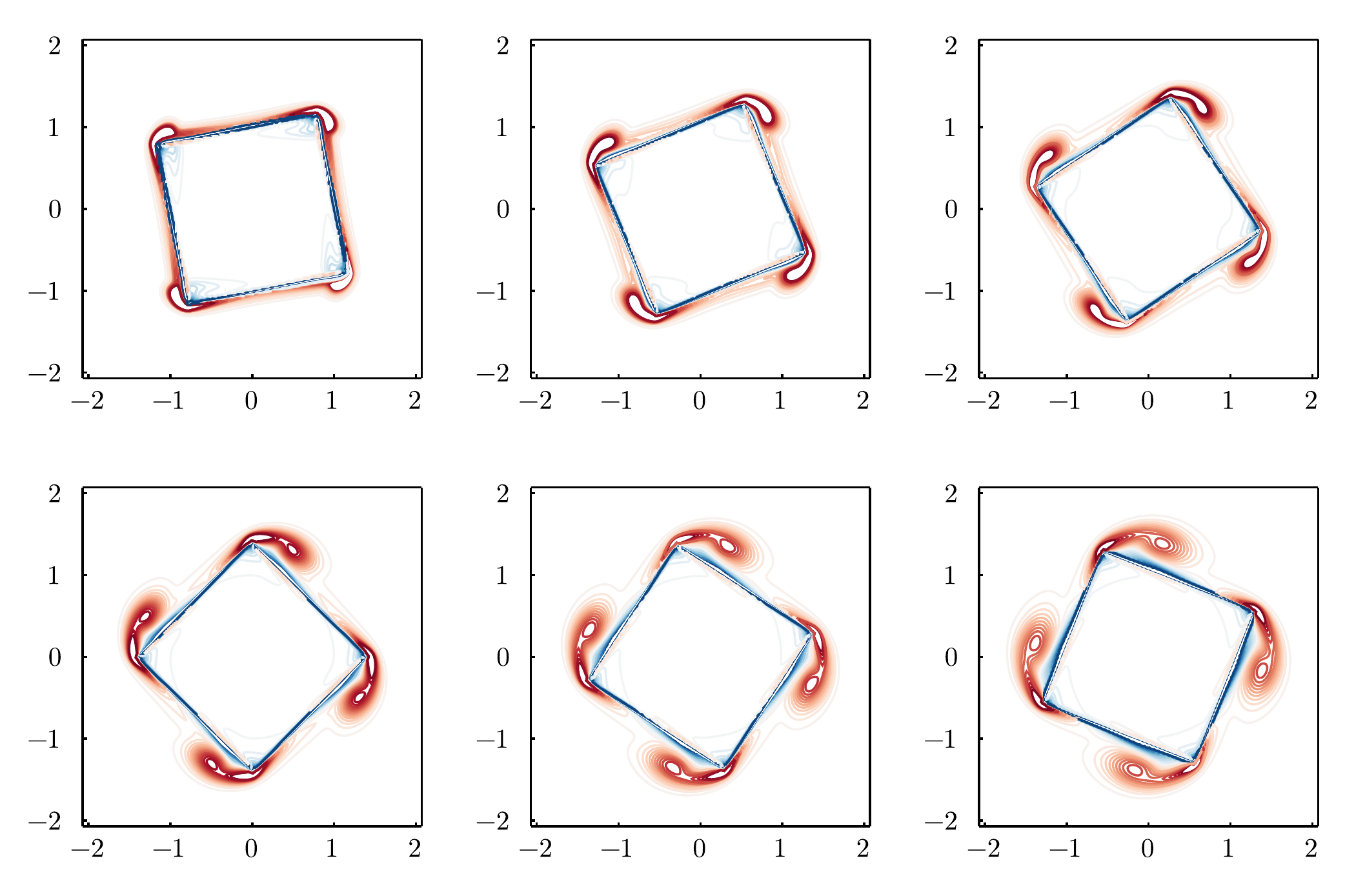}
\caption{Snapshots of vorticity contours at $\Omega t = 0.2$ through $1.2$ (left to right, then top to bottom) for impulsive rotation of square using external IBPM. Vorticity contours between $\omega/\Omega = -10$ and $10$ are shown.}\label{fig:square-il}
\end{figure}

\begin{figure}[bt]
\centering
\includegraphics[width=\textwidth]{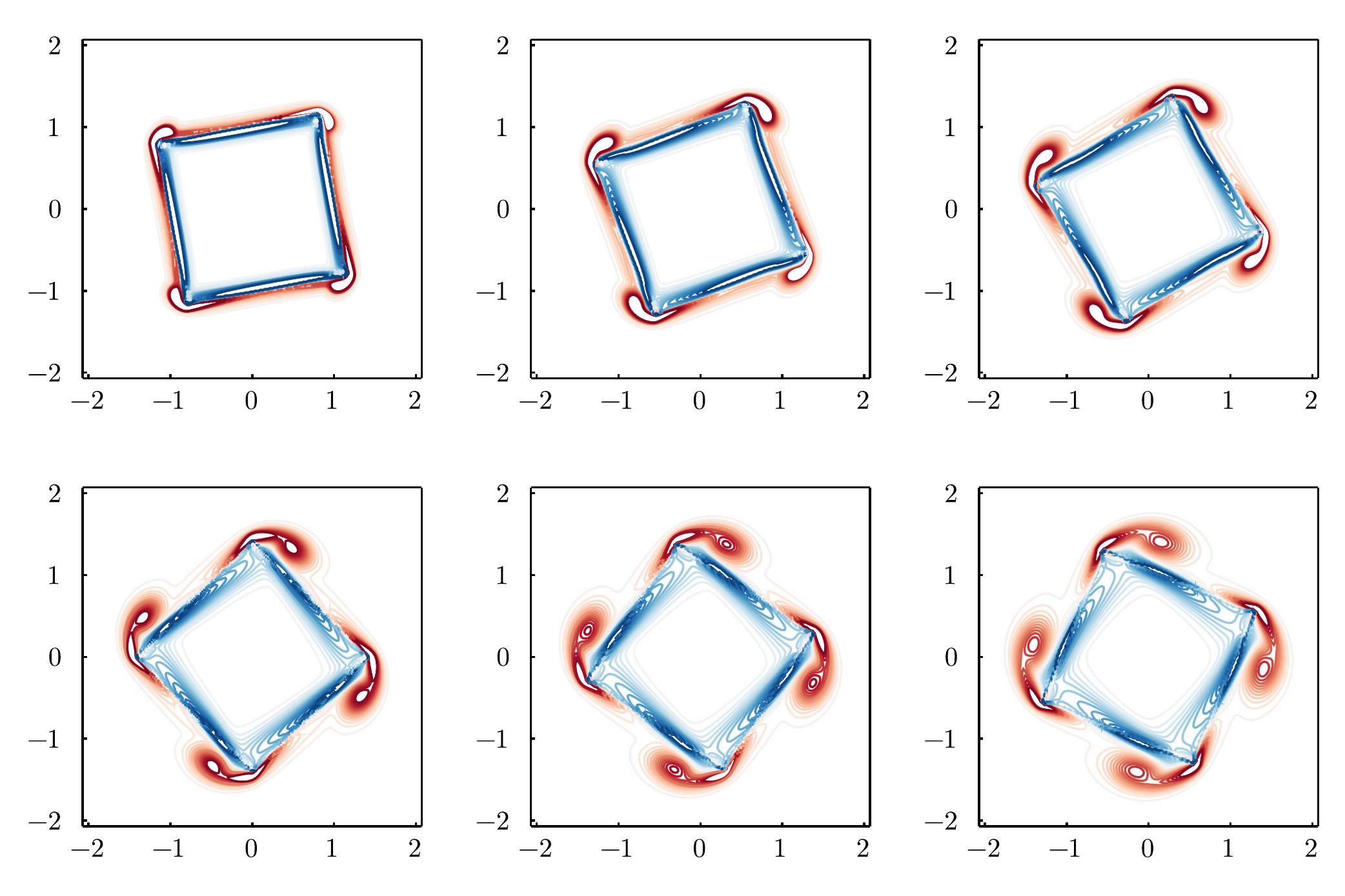}
\caption{Snapshots of vorticity contours at $\Omega t = 0.2$ through $1.2$ (left to right, then top to bottom) for impulsive rotation of square using original IBPM. Vorticity contours between $\omega/\Omega = -10$ and $10$ are shown.}\label{fig:square-ibpm}
\end{figure}

In the next example, we solve for the flow exterior to an impulsively-rotating square of side length $L$, at Reynolds number $\Omega L^{2}/\nu = 400$. This case serves as a test of the discrete immersed layers to resolve the flow in the vicinity of a sharp corner. We use a grid spacing of $\dx/L = 0.02$ and 260 forcing points distributed uniformly about the square's perimeter; the time step size is set to $\nu\Delta t/\Delta x^{2} = 0.25$ (or $\Omega \Delta t = 0.01$). Figure~\ref{fig:square-il} depicts the resulting vorticity using the current method. A small amount of vorticity is observed in the interior regions, but this leakage remains constant as the exterior vorticity develops due to flow separation from the corners. In contrast, the original IBPM \cite{taira2007,liskacolonius17} generates an interior flow, as expected, shown in Figure~\ref{fig:square-ibpm}.

\section{Conclusions}

In this work, we have developed a method of immersed layers, an extended form of the immersed boundary method (IBM) that enables us to distinguish the sides of the immersed surface and to impose different conditions on each. By defining a discrete version of a masking (Heaviside) function that acts as a natural companion to the discrete Dirac delta function in the IBM, we have obtained a useful set of discrete identities that mimic continuous ones that are known from the theory of generalized functions. With these operators and identities, we have defined a discretely-masked field on a Cartesian grid and shown that finite differencing operators that act on this field naturally give rise to immersed layers, containing jumps in the grid field on either side of the immersed boundary. In several prototype partial differential equations---the Poisson equation, the convection-diffusion equation, and incompressible Navier--Stokes---we have shown that the differencing operators on the masked fields introduce immersed layer terms into these equations, including the regularized force term familiar from immersed boundary methods as well as additional terms containing jumps in the grid fields. We identify these field jumps with surface quantities, some of which are known (e.g., prescribed velocities on either side) and others of which are solved for (e.g., surface tractions). The new method achieves the same accuracy and flexibility as the immersed boundary method (and particularly, the immersed boundary projection method of Taira and Colonius \cite{taira2007}). However, with its freedom to set boundary conditions on each side of the boundary, the new method readily enables us to avoid the creation of a superfluous solution in the opposite region. This ensures, in the particular case of incompressible flow with velocity boundary conditions, that the computed Lagrange multipliers represent the local traction exerted by the fluid (plus a prescribed momentum flux) rather than an indistinct mix of tractions from the intended and superfluous sides. It is straightforward to enforce traction-type boundary conditions in the method, as well, and this is the subject of current work.

Though we have only demonstrated the new method on a sampling of representative two-dimensional problems, there is no aspect of the method that limits it to two dimensions. The essence of the method is the regularization of surface data and unit normals onto the grid. All of the essential operators are already standard to the IBM and finite difference methods and extend in the usual manner to three dimensions; unit normal vectors are readily definable on discrete surfaces.  The method can be used for multiple disjoint surfaces, as well, without any modification of the underlying operators' structure. We have focused on Dirichlet boundary conditions in this paper, but the immersed layers give access to boundary values of any natural type. For example, conditions could alternatively be placed on surface fluxes, and the solution obtained as for Neumann problems in partial differential equations \cite{stakgold2000boundary}. By such means, one could in principle use the immersed layers to prescribe inflow and outflow conditions in internal flows, an approach we are pursuing in current work.

Finally, we observe that the masked fields and the associated extended partial differential equations that we have used in this work express a general principle of the immersion of surfaces into Euclidean space. Though we have applied this principle in a particular manner in this work to develop a numerical method and solve problems, the framework also creates an opportunity to analyze and compare most other immersed boundary methods (in the larger class discussed in Section~\ref{sec:intro}) in a common setting. It would be interesting, for example, to attempt to express incompressible flow methods that directly introduce the boundary conditions into the finite difference operators, e.g., sharp interface \cite{mittal2008versatile}, immersed interface \cite{leelev:1j}, and ghost fluid methods \cite{fedkiw1999non}, in this context. We leave this attempt for future work.

\section*{Acknowledgments}
Support by the U.S. Air Force Office of Scientific Research FA9550-18-1-0440 is gratefully acknowledged.


\appendix
\section{A brief review of generalized functions}
\label{sec:genfun}

In this section, we provide an overview of the necessary definitions, properties, and identities of generalized functions. More details can be found in various sources, including Farassat \citep{farassat96}

\subsection{The basic definitions}
The Heaviside function $\heavi(y)$ is defined as
\begin{equation}
\label{eq:heavi}
\heavi(y) = \left\{ \begin{array}{ll} 1, & y > 0, \\ 1/2, & y =0, \\ 0, & y < 0. \end{array} \right.
\end{equation}
It is clear from this definition that $\heavi(-y) = 1 - \heavi(y)$. The defining feature of this function is that it allows us to extend an integral over a semi-infinite interval into an infinite range. Consider an integrable (and differentiable) function $f(y)$. Then
\begin{equation}
\int_{-\infty}^{x} f(y)\,\mathrm{d}y = \int_{-\infty}^{\infty} H(x-y) f(y)\,\mathrm{d}y
\end{equation}

Suppose we take the derivative of both sides with respect to $x$. Then we get
\begin{equation}
\label{eq:sift}
f(x) = \int_{-\infty}^{\infty} H'(x-y) f(y)\,\mathrm{d}y,
\end{equation}
where $()'$ denotes the derivative of a function with respect to its argument. This first derivative of the Heaviside function is the Dirac delta function, $\dirac(y) = \heavi'(y)$, whose defining property is expressed in equation (\ref{eq:sift}), rewritten here:
\begin{equation}
\label{eq:siftdirac}
f(x) = \int_{-\infty}^{\infty} \dirac(x-y) f(y)\,\mathrm{d}y.
\end{equation}
Note that $\dirac$ is an even function of its argument: $\dirac(-x) = \dirac(x)$. Also, if we take $f$ to be simply equal to 1 in equation (\ref{eq:siftdirac}), then the integral of $\dirac$ over the real axis is unity:
\begin{equation}
\label{eq:siftdirac2}
1 = \int_{-\infty}^{\infty} \dirac(y)\,\mathrm{d}y.
\end{equation}

By its definition as the derivative of the Heaviside function, we must have that
\begin{equation}
\dirac(y) = \left\{ \begin{array}{ll} \infty, & y = 0,  \\ 0, & y \neq 0. \end{array} \right.
\end{equation}

We can also define a multi-dimensional version of $\dirac$. For $\x = (x_{1},x_{2},\ldots,x_{d}) \in \R^{d}$, where $d$ is the spatial dimension (1, 2, or 3), then
\begin{equation}
\dirac(\x) = \dirac(x_{1})\dirac(x_{2})\cdots\dirac(x_{d}),
\end{equation}
and all properties of this multi-dimensional version follow from the properties of the one-dimensional form. In particular, its integral remains unity:
\begin{equation}
\label{eq:siftdiracd}
\int_{\R^{d}} \dirac(\x)\,\dvol = 1.
\end{equation}

Now let us take a derivative of equation (\ref{eq:siftdirac}). We get
\begin{equation}
\label{eq:ddirac}
f'(x) = \int_{-\infty}^{\infty} \dirac'(x-y) f(y)\,\mathrm{d}y.
\end{equation}
Thus, the derivative of $\dirac$, when integrated with a differentiable function, picks off the derivative of that function. Note that $\dirac'(-y) = -\dirac'(y)$.
 
\subsection{Generalized functions applied to an indicator function}
\label{sec:genindic}
The definitions of generalized functions provide a useful means of distinguishing the left and right sides of the zero value of the argument. This property forms the foundation for a useful set of tools when the functions are applied to an {\em indicator function} (sometimes called a characteristic function) whose zero level set is designed to implicitly identify, say, the surface of a body in $\R^{d}$. In that spirit, let us consider a differentiable function $\indic(\x)$, for $\x \in \R^{d}$, whose level set $\indic=0$ separates a region in which $\indic < 0$ (which we will refer to as the {\em interior}, and which we will assume to be of finite size) from another region in which $\indic > 0$ (the {\em exterior}, which we will assume to extend to infinity). It should be noted that the interior may comprise more than one disjointed region without consequence; these definitions are illustrated in Figure~\ref{fig:indicatorfcn}. The gradient $\grad \indic$, when evaluated on the surface $\indic = 0$, is proportional to the unit normal, $\nrm$, directed toward the exterior. In fact, we can always define $\indic$ so that $|\nabla \indic| = 1$ on $\indic=0$, ensuring that $\grad \indic$ is identically the unit normal:
\begin{equation}
\grad\indic = \nrm.
\end{equation}

\subsubsection{Basic composite functions} By definition (\ref{eq:heavi}), the composite function $\heavi(\indic(\x))$ is equal to 1 when $\x$ lies in the exterior, 0 for $\x$ in the interior, and $1/2$ for $\x$ on the level set $\indic=0$; the composite $\heavi(-\indic(\x))$ switches the regions to which it assigns 1 and 0. With this behavior, these composite functions inherit a very useful {\em masking} property. For example, we can write integrals of a function $f$ over the exterior or interior alternatively as integrals of the product $f$ with these composite Heaviside functions over the entire space $\R^{d}$:
\begin{equation}
\label{eq:heavig}
\int_{\indic>0} f(\x) \,\dvol = \int_{\R^{d}} f(\x) \heavi(\indic(\x))\,\dvol,\qquad \int_{\indic<0} f(\x) \,\dvol = \int_{\R^{d}} f(\x) \heavi(-\indic(\x))\,\dvol.
\end{equation}
For shorthand, we will define $\mask{f} = f \heavi(\indic)$ and refer to this as the masked version of $f$.

The composite function $\dirac(\indic(\x))$ has the similarly useful ability to rewrite a surface integral over the level set $\indic=0$ as a volume integral over $\R^{d}$. Suppose this surface is parameterized by surface coordinate(s) $\surfc$, so that points on this surface when embedded in $\R^{d}$ can be described by $\x = \X(\surfc)$. Then the surface integral can be written as
\begin{equation}
\label{eq:diracg}
\int_{\indic=0} f(\X(\surfc))\,\dsurf = \int_{\R^{d}} f(\x) \dirac(\indic(\x))\,\dvol.
\end{equation}
It follows, then, that the product $f(\x) \dirac(\indic(\x))$ can be written alternatively as
\begin{equation}
\label{eq:diracg2}
f(\x) \dirac(\indic(\x)) = \int_{\indic=0} f(\X(\surfc)) \dirac(\x - \X(\surfc))\,\dsurf,
\end{equation}
for if one integrates both sides of this equation over $\R^{d}$ and uses equations (\ref{eq:siftdiracd}) and (\ref{eq:diracg}), then both sides reduce to an integral of $f$ over the surface $\indic=0$. It should be observed that only the values of $f$ on the surface are actually invoked in this product. In fact, the operation can be applied to a function, $F(\surfc)$, defined only on the surface, so that it is extended to a definition over all of $\R^{d}$:
\begin{equation}
\label{eq:diracg3}
F\dirac(\indic(\x)) = \int_{\indic=0} F(\surfc) \dirac(\x - \X(\surfc))\,\dsurf.
\end{equation}
In other words, the product $F\dirac(\indic)$ embeds (or {\em immerses}) a surface function $F$ into $\R^{d}$. If one takes the trivial surface function $F = 1$, then equation (\ref{eq:diracg2}) becomes an identity for the immersion function $\dirac(\indic)$: 
\begin{equation}
\label{eq:diracg2unity}
\dirac(\indic(\x)) = \int_{\indic=0} \dirac(\x - \X(\surfc))\,\dsurf,
\end{equation}
for all $\x \in \R^{d}$. Note that $\dirac(-\indic) = \dirac(\indic)$.

Suppose we consider the product of the embedded surface function $F \dirac(\indic)$ with another integrable function, $g(\x)$, over all of $\R^{d}$. Then, by our definitions thus far,
\begin{equation}
\int_{\R^{d}} g(\x) F \dirac(\indic(\x))\,\dvol = \int_{\R^{d}} g(\x) \int_{\indic=0} F(\surfc) \dirac(\x - \X(\surfc))\,\dsurf\,\dvol.
\end{equation}
The integrals over $\R^{d}$ and the surface $\indic=0$ can be swapped in the last expression, so that we have
\begin{equation}
\int_{\indic=0} F(\surfc) \int_{\R^{d}} g(\x) \dirac(\X(\surfc) - \x)\,\dvol \,\dsurf.
\end{equation}
But, by property (\ref{eq:siftdirac}), the inner integral {\em restricts} the function $g(\x)$ to the surface $\indic=0$,
\begin{equation}
\label{eq:restrict}
\int_{\R^{d}} g(\x) \dirac(\X(\surfc) - \x)\,\dvol = g(\X(\surfc)).
\end{equation}
Overall, we have shown that
\begin{equation}
\int_{\R^{d}} g(\x) F \dirac(\indic(\x))\,\dvol  = \int_{\indic=0} g(\X(\surfc)) F(\surfc) \,\dsurf.
\end{equation}
Because each of these integrals represents a standard inner product for functions on the respective domains, we can think of the restriction operation (\ref{eq:restrict}) as the transpose of the immersion operation (\ref{eq:diracg3}). We will denote this restriction in shorthand as
\begin{equation}
\label{eq:diracT}
 g(\x) \dirac^{T}(\indic(\x)) = \int_{\R^{d}} g(\x) \dirac(\X(\surfc) - \x)\,\dvol = g(\X(\surfc)).
\end{equation}

\subsubsection{Masked functions and their derivatives} \label{sec:maskfcn}
Before we continue further, it is important to anticipate the role of the spatial gradient of $\heavi$, which introduces the directionality of the surface $\indic=0$:
\begin{equation}
\label{eq:gradheavi}
\grad\heavi(\pm\indic(\x)) = \pm \heavi'(\indic(\x))\grad\indic = \pm \dirac(\indic(\x)) \nrm(\x),
\end{equation}
or simply, $\grad\heavi(\pm\indic) = \pm\dirac(\indic) \nrm$.

If the surface is in motion, then it is also important to develop the time derivative of $\heavi$. In such a case, the indicator function and surface points are both functions of time, $t$, and the velocity of the surface is given by $\partial\surfpt/\partial t$. For shorthand, we will denote this velocity by $\surfvel$. By definition, the indicator function is invariant on the surface defined by its level set $\indic=0$; thus, its time derivative is given by
\begin{equation}
\ddp{\indic}{t} = -\surfvel\cdot\grad\indic = -\surfvel\cdot\nrm.
\end{equation}
The time derivative of the masking function $\heavi$ follows from the chain rule,
\begin{equation}
\ddp{\heavi(\pm\indic(\x,t))}{t} = \pm\heavi'(\pm\indic(\x,t)) \ddp{\indic}{t},
\end{equation}
and it is easy to see that
\begin{equation}
\label{eq:ddtheavi}
\ddp{\heavi(\pm\indic)}{t} = \mp\surfvel\cdot\dirac(\indic)\nrm.
\end{equation}

With these identities on $\heavi$ established, let us consider the gradient and time derivative of the masked function $\mask{f} = f \heavi(\indic)$,
\begin{equation}
\label{eq:dmaskf}
\grad \mask{f} = \heavi(\indic) \grad f + f \dirac(\indic) \nrm
\end{equation}
and
\begin{equation}
\label{eq:ddtmaskf}
\ddp{\mask{f}}{t} = \heavi(\indic) \ddp{f}{t} - f \surfvel\cdot\dirac(\indic) \nrm.
\end{equation}
It should be noted that the gradient of $f$ on the right-hand side of (\ref{eq:dmaskf}) is to be interpreted as a continuous derivative over all of the exterior, including the limit as the level set $\indic=0$ is approached from within this region. Furthermore, the material derivative of $\mask{f}$, in a velocity field $\fluidvel$, is easily shown to be
\begin{equation}
\label{eq:DDtmaskf}
\ddp{\mask{f}}{t} + \fluidvel\cdot\grad\mask{f} = \heavi(\indic) \left( \ddp{f}{t} + \fluidvel\cdot\grad f \right) + f (\fluidvel-\surfvel)\cdot \dirac(\indic) \nrm.
\end{equation}

Let us now integrate the identity (\ref{eq:dmaskf}) over the space $\R^{d}$.  Using the identities (\ref{eq:heavig}) and (\ref{eq:diracg}), we get
\begin{equation}
\int_{\R^{d}} \grad \mask{f}\,\dvol = \int_{\indic > 0}\grad f \,\dvol + \int_{\indic=0} f\nrm\,\dsurf.
\end{equation}
We have already tacitly assumed that the function $f$ is integrable. Let us now further restrict $f$ so that $f \rightarrow 0$ as $|\x| \rightarrow \infty$. Then, if we apply the divergence theorem to the integral on the left-hand side, we find that it vanishes, and we recover the divergence theorem over the exterior (remembering that $\nrm$ is directed into this region):
\begin{equation}
\int_{\indic > 0}\grad f \,\dvol = - \int_{\indic=0} f\nrm\,\dsurf.
\end{equation}
In other words, by applying differential operations to the masked function $\mask{f} = f \heavi(\indic)$ and considering its domain over all of $\R^{d}$, we automatically obtain the expected properties on the immersed surface $\indic = 0$. 

We can naturally extend the masking concept to a form that combines separate definitions of the function in the interior and exterior. For example, suppose that
\begin{equation}
\label{eq:fmulti}
f(\x) = \left\{ \begin{array}{ll} f^{+}(\x), & \x \in \indic > 0, \\ \frac{1}{2} (f^{+}(\x)+f^{-}(\x)), & \x \in \indic = 0, \\ f^{-}(\x), & \x \in \indic < 0. \end{array} \right.
\end{equation}
This piecewise function can be easily expressed in terms of the exterior and interior masks,
\begin{equation}
\label{eq:maskf}
\mask{f} \equiv f^{+} \heavi(\indic) + f^{-} \heavi(-\indic).
\end{equation}
It is easy to verify, from the property of the Heaviside function, that when the masked function $\mask{f}$ is restricted to the surface $\indic=0$, we recover the average of the exterior and interior values on the surface.
\begin{equation}
\mask{f} \dirac(\indic) = \frac{1}{2} (f^{+} + f^{-}) \dirac(\indic).
\end{equation}

Applying the gradient on this form of the masked function, we get
\begin{equation}
\label{eq:gradf0}
\grad \mask{f} = \heavi(\indic) \grad f^{+} + \heavi(-\indic) \grad f^{-} + (f^{+}-f^{-}) \dirac(\indic) \nrm.
\end{equation}
 Thus, the gradient of the masked function naturally accounts for the jump in $f$ on the level set $\indic=0$. As a shorthand, let us define the mask of the gradient to be
 \begin{equation}
 \mask{\grad f} \equiv \heavi(\indic) \grad f^{+} + \heavi(-\indic) \grad f^{-}, 
 \end{equation}
 so we can succinctly write (\ref{eq:gradf0}) as
 \begin{equation}
\label{eq:gradf}
\grad \mask{f} = \mask{\grad f}+ (f^{+}-f^{-}) \dirac(\indic) \nrm.
\end{equation}
 For general spatiotemporal differential operator $D$, we extend this definition of the mask:
 \begin{equation}
 \label{eq:maskDf}
 \mask{D f} \equiv \heavi(\indic) D f^{+} + \heavi(-\indic) D f^{-}.
 \end{equation}
 We will use this notation in several other identities below.
 
 As in equation (\ref{eq:dmaskf}), each of the gradients on the right-hand side should be interpreted within its respective region, continuous in the limit as the level set $\indic=0$ is approached in this region. As a special case, if we let $f^{+} = 0$ and $f^{-} = 1$---so that the gradients of both vanish---and then integrate the equation over $\R^{d}$, then we recover the basic identity on closed surfaces,
 \begin{equation}
 \label{eq:normint}
 \int_{\R^{d}} \dirac(\indic) \nrm\,\dvol = \int_{\indic=0} \nrm\,\dsurf = 0.
 \end{equation}

 The time derivative can also be applied to the masked function $\mask{f}$ in (\ref{eq:maskf}), extending the result of (\ref{eq:ddtmaskf}):
 \begin{equation}
\label{eq:ddtf}
\ddp{\mask{f}}{t} = \mask{ \ddp{f}{t}} - (f^{+}-f^{-}) \surfvel\cdot\dirac(\indic) \nrm;
\end{equation}
similarly, we can show that the material derivative of $\mask{f}$, with a masked velocity field $\mask{\fluidvel}$ defined in the same manner as (\ref{eq:maskf}), is:
  \begin{equation}
\label{eq:DDtf}
\ddp{\mask{f}}{t} + \mask{\fluidvel}\cdot\grad\mask{f} =  \mask{ \ddp{f}{t} + \fluidvel\cdot\grad f} + (f^{+}-f^{-}) (\mask{\fluidvel} - \surfvel)\cdot\dirac(\indic) \nrm,
\end{equation}
where $\fluidvel$ inside the mask operator on the right-hand side is set to the velocity in each respective region. Each of the two derivatives of the masked field $\mask{f}$ contains a term proportional to $\dirac(\indic)\nrm$. In particular, the final term in (\ref{eq:DDtf}) serves as a flux of the jump $f^{+}-f^{-}$ across the interface $\indic=0$. This flux is absent if the interface's normal velocity matches that of $\mask{\fluidvel}$, which itself is equal to the average of the normal velocities of $\fluidvel^{+}$ and $\fluidvel^{-}$.

If $\vecfunc$ is a vector field whose components have the same degree of integrability and differentiability that we have supposed for $f$, then we can obtain similar properties for vector differential operations when we mask $\vecfunc$, e.g.,
\begin{equation}
\label{eq:divf}
\div \mask{\vecfunc} = \mask{\div\vecfunc} + \nrm\cdot(\vecfunc^{+} - \vecfunc^{-}) \dirac(\indic)
\end{equation}
and
\begin{equation}
\label{eq:curlf}
\curl \mask{\vecfunc} = \mask{\curl\vecfunc} + \nrm\times(\vecfunc^{+} - \vecfunc^{-}) \dirac(\indic).
\end{equation}
Similarly, if $\tensfunc$ is a rank-2 tensor field, also with integrable and differentiable components, then the divergence of $\mask{\tensfunc}$ is
\begin{equation}
\div \mask{\tensfunc} = \mask{\div\tensfunc} + \nrm\cdot(\tensfunc^{+} - \tensfunc^{-}) \dirac(\indic).
\end{equation}
The integrals of these identities over $\R^{d}$ would lead to corresponding forms of the divergence theorem in the respective regions.

We can combine the differential operations of the masked field presented in this section and obtain useful results. For example, it can be shown that the divergence of the curl of the masked vector field vanishes identically, as does the curl of the gradient of a masked scalar field. The divergence of the gradient of a masked field is used in Section~\ref{sec:poisson} to develop an extended form of the Poisson equation.

\subsection{Green's function solution of Poisson equation}
\label{sec:green}

Equation (\ref{eq:green}) expresses the effect of a source centered at the origin. The solution is invariant to translation---that is, $G(\x-\y)$ is the solution of the equation for right-hand side $-\dirac(\x-\y)$. Furthermore, it should also be noted that, like $\dirac$, the Green's function is even with respect to its argument: $G(\x-\y) = G(\y-\x)$; in other words, a source and target point exert the same influence on each other. However, the gradient is an odd function, $\grad_{\y} G(\x-\y) = -\grad_{\x} G(\x-\y) = -\grad G(\x - \y)$.

Now consider the identity
\begin{align}
\nabla \cdot \left[ G(\x-\y) \grad \mask{\varphi}(\x) -\mask{\varphi}(\x) \grad G(\x-\y) \right] &= \nonumber \\
 & \hspace{-2cm} G(\x-\y) \lap \mask{\varphi}(\x) -\mask{\varphi}(\x) \lap G(\x-\y),
\end{align}
integrated with respect to $\x$ over $\R^{d}$. The left-hand side of the integrated equation vanishes by virtue of the divergence theorem and the fact that $\mask{\varphi}$ vanishes at infinity. The Laplacians in the expression on the right-hand side can be replaced with (\ref{eq:lapphi}) and (\ref{eq:green}), respectively. Using the property (\ref{eq:siftdirac}) of the function $\dirac$, we get
\begin{align}
\label{eq:phisoln1}
\mask{\varphi}(\y) &= -\int_{\R^{d}} G(\x-\y) \mask{q}(\x) \,\dvol  - \int_{\R^{d}}  G(\x-\y) \nrm\cdot(\grad\varphi^{+} - \grad\varphi^{-}) \dirac(\indic)\,\dvol \nonumber \\
&  \hspace{2cm} - \int_{\R^{d}} G(\x-\y) \div \left( (\varphi^{+}-\varphi^{-}) \dirac(\indic) \nrm \right)\,\dvol.
\end{align}
The third integral can be rewritten in a somewhat more helpful form by first integrating by parts, based on the identity
\begin{align}
G(\x-\y) \div \left( (\varphi^{+}-\varphi^{-}) \dirac(\indic) \nrm \right) &= \div \left( G(\x-\y) (\varphi^{+}-\varphi^{-}) \dirac(\indic) \nrm \right) \nonumber \\ 
& \hspace{2cm} - \grad G(\x - \y) \cdot (\varphi^{+}-\varphi^{-}) \dirac(\indic) \nrm.
\end{align}
The first term on the right-hand side vanishes when integrated over $\R^{d}$, by virtue of the divergence theorem and the fact that $\dirac(\indic)$ is zero everywhere except on $\indic=0$. The last term on the right-hand side can be rewritten as $- \grad G(\x - \y) \cdot (\varphi^{+}-\varphi^{-}) \dirac(\indic) \nrm = \grad_{\y} G(\x - \y) \cdot (\varphi^{+}-\varphi^{-}) \dirac(\indic) \nrm$. But since only the Green's function depends on $\y$, this can be written instead as $\nabla_{\y} \cdot \left( G(\x - \y) (\varphi^{+}-\varphi^{-}) \dirac(\indic) \nrm \right)$. Thus, when this is introduced into the full expression (\ref{eq:phisoln1}) and integrated with respect to $\x$ over $\R^{d}$, we obtain the formal solution of the generalized Poisson equation,
\begin{align}
\label{eq:phisolnapp}
\mask{\varphi}(\y) &= -\int_{\R^{d}} G(\x-\y) \mask{q}(\x) \,\dvol  - \int_{\R^{d}}  G(\x-\y) \nrm\cdot(\grad\varphi^{+} - \grad\varphi^{-}) \dirac(\indic)\,\dvol \nonumber \\
&  \hspace{5cm} - \nabla_{\y} \cdot \int_{\R^{d}} G(\x - \y) (\varphi^{+}-\varphi^{-}) \dirac(\indic) \nrm\,\dvol.
\end{align}

\section{Discrete operators}
\label{sec:discrete}

In this section, we present the details of the grid and immersed point spaces that are used in the paper. Many of these operators mimic those used in the previous section. The fields are expressed on a staggered Cartesian grid of uniform spacing $\dx$ and infinite extent. 

\begin{figure}[t]
\begin{center}
\includegraphics{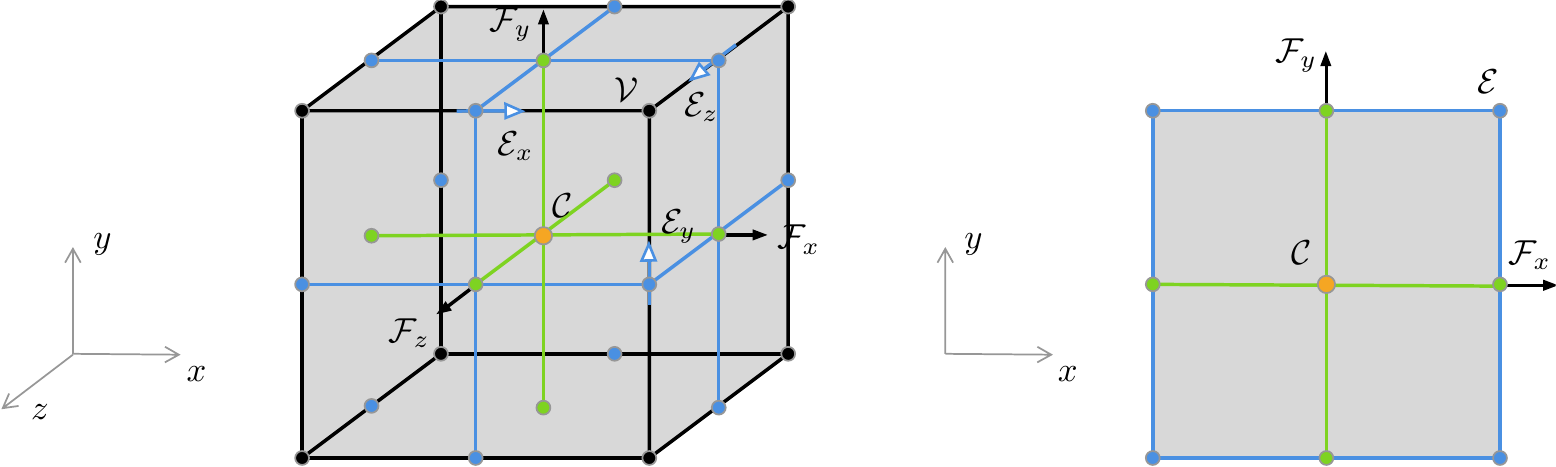}
\caption{Schematic of a grid cell $\idex$ in 3-d (left) and 2-d (right).  Locations holding components of the grid vector spaces are denoted by circles. Orange: cell centers, $\centers$; Green: cell faces, $\faces$; Blue: cell edges, $\edges$; Black: cell vertices, $\vertices$. The labeled locations each share the same index $\idex$. The colored lines join the points invoked to calculate differences in $\divgrid$ and $\curlgrid$ or interpolations in $\interpgrid{\faces}{\centers}$ and $\interpgrid{\tensors}{\faces}$.}
\label{fig:unitcell}
\end{center}
\end{figure}

\subsection{Grid spaces and operators} On the grid itself, we will use mimetic finite differencing operators \citep{coloniustaira08:1j,liskacolonius17} that are endowed with many of the same properties possessed by their continuous counterparts. These include discrete versions of the divergence $\divgrid$, curl $\curlgrid$, gradient $\gradgrid$, and Laplacian $\lapgrid$ operators, each obtained by second-order-accurate central differencing between the nodes, edges, and faces of the staggered grid. Note that, by our definition in this paper, each of these operators is scaled by the physical grid spacing so that it approximates the corresponding continuous operator to second order. Importantly, we will make extensive use of the so-called lattice Green's function, $\lgf$, on this grid, which allows us to formally invert the discrete Laplacian on a uniform grid of infinite extent and form discrete analogs of the results from Green's theorem. The combination of these tools will allow us to construct a discrete Heaviside function and several other useful tools---particularly, a discrete masking function.

To aid the discussion that follows, let $\idex = (i_{1},i_{2},\ldots,i_{d})$ denote a multi-index that uniquely describes a cell on the grid, illustrated in Figure~\ref{fig:unitcell}. As alluded to above, there are several different spaces associated with a staggered grid: cell centers and vertices, which hold scalar-valued data, and cell faces and edges, which contain vector-valued data. These latter two spaces are a Cartesian product of the spaces holding the vector components: $\faces = \faces_{x} \times \faces_{y} \times \faces_{z}$ and $\edges = \edges_{x} \times \edges_{y} \times \edges_{z}$. The space of cell centers, $\centers$, typically holds the pressure and scalar potential fields; cell faces, $\faces$, contains the velocity field; and cell edges, $\edges$, the vorticity field; the cell vertices have limited use, but would hold the divergence of edge data. In two dimensions, the grid contains only the cell centers $\centers$, the faces $\faces_{x}$ and $\faces_{y}$, and the edge aligned with the $z$ direction, which we abbreviate to $\edges$; the vertices disappear entirely. We will also have need to represent tensors on the grid, particularly those that arise from the gradient of a vector field in $\faces$ (or a Cartesian product of two such vectors). This space, denoted by $\tensors$, is composed entirely of the component spaces $\centers$ and $\edges$:
\begin{equation}
\label{eq:tensors}
\tensors = \begin{bmatrix} \centers & \edges_{z} & \edges_{y} \\ \edges_{z} & \centers & \edges_{x} \\ \edges_{y} & \edges_{x} & \centers \end{bmatrix}.
\end{equation}
(In two dimensions, the space contains only the upper left $2 \times 2$ block.)

The data associated with a given discretized quantity belonging to any one of these spaces on the grid comprises a vector, which we denote in lower case. For example, a scalar potential could be denoted by $\gridvar{\solnvec} \in \centers$. Because the grid is unbounded in all directions, it is safe to assume that every cell with multi-index $\idex$ has a single entry in each of the grid spaces. Each component of the vector is uniquely associated with a multi-index $\idex$. We will denote a component of any vector---say, $\gridvar{\solnvec}$---by $\gridcomp{\solnvec}{\idex}$. For a vector of face or edge data, we will distinguish the $d$ components on each face or edge by a numerical superscript, e.g., for $\gridvar{\vecvec}\in \faces$, $\vgridcomp{\vecvec}{\idex}{1}$ represents the component on on the face $\faces_{x}$ with index $\idex$. As a convention, we will assume that the $d$ faces sharing the same multi-index with the cell center comprise the right/top/front faces of the cell, and that the edge(s) with this multi-index form the intersections of these faces with one another, as shown in Figure~\ref{fig:unitcell}. Occasionally, we must compute element-by-element products between two vectors of the same space, returning another vector in the same space. We use the symbol $\had$ to denote such a product. For vectors in $\faces$, the product is carried out separately for each of the $d$ components. 

The grid spaces each have a natural inner product based on a sum over all of the grid points; to ensure that it approximates an integral over $\R^{d}$, we multiply the sum by the cell volume, $\dx^{d}$. We will label each inner product with the associated space, e.g., for $\gridvar{f}_{1},\gridvar{f}_{2} \in \centers$, the inner product of cell-centered data is
\begin{equation}
\ipcenters {\gridvar{f}_{1}}{\gridvar{f}_{2}} = \dx^{d} \sum_{\idex} \gridcomp{f_{1}}{\idex} \gridcomp{f_{2}}{\idex}.
\end{equation}
If we let $\onesgrid \in \centers$ denote a vector of grid data equal to 1 at all cell centers, and $\gridvar{f} \in \centers$ a sampling of the continuous (and integrable) function $f(\x)$ on these points, then
\begin{equation}
\ipcenters {\onesgrid}{\gridvar{f}} = \dx^{d} \sum_{\idex} \gridcomp{f}{\idex}
\end{equation}
 represents an approximation of the integral of $f(\x)$ over $\R^{d}$.
 
Another grid vector that will come in use later is $\unitgrid_{\idex} \in \centers$, which denotes a set of data equal to $1$ at the cell center with index $\idex$ and equal to $0$ at every other point. The collection of vectors $\unitgrid_{\idex}$ for all possible $\idex$---a countably infinite set on this unbounded grid---forms a unit basis for $\centers$. That is, we can write any $\gridvar{f} \in \centers$ as
\begin{equation}
\label{eq:celldecomp}
\gridvar{f}  = \sum_{\idex} \gridcomp{f}{\idex} \unitgrid_{\idex}.
\end{equation}

A similar inner product can be defined in $\faces$, but must now sum over the $d$ faces that share the same index with a cell. For $\gridvar{\vecvec}_{1},\gridvar{\vecvec}_{2} \in \faces$,
 \begin{equation}
 \ipfaces{\gridvar{\vecvec}_{1}}{\gridvar{\vecvec}_{2}} = \dx^{d} \sum_{k=1}^{d} \sum_{\idex}  \vgridcomp{\vecvec_{1}}{\idex}{k} \vgridcomp{\vecvec_{2}}{\idex}{k}.
 \end{equation}
The special vectors $\onesgrid$ and $\unitgrid_{\idex}$ also serve important roles in $\faces$ as they do in $\centers$, but only when distinguished with a particular direction, indicated by a superscript. For example, we will let $\onesgrid^{(2)} \in \faces$ denote the vector uniformly equal to 1 on all faces in $\faces_{y}$. Similarly, $\unitgrid_{\idex}^{(3)} \in \faces$ is equal to $1$ only at the face $\idex$ in $\faces_{z}$.

The discrete gradient and divergence operators map data between the cell centers and faces: $\gradgrid: \centers \mapsto \faces$ and $\divgrid: \faces \mapsto \centers$. With the inner products we have defined on $\centers$ and $\faces$, it can be shown that the discrete divergence $\divgrid$ and gradient $\gradgrid$ are (negative) adjoints of one another. For any $\gridvar{\solnvec} \in \centers$ and $\gridvar{\vecvec} \in \faces$,
\begin{equation}
\label{eq:divTgrad}
\ipcenters {\gridvar{\solnvec}}{\divgrid\gridvar{\vecvec}} = -\ipfaces {\gradgrid\gridvar{\solnvec}}{\gridvar{\vecvec}}.
\end{equation}
We will also need the discrete curl $\curlgrid: \edges \mapsto \faces$ and its adjoint $\rotgrid: \faces \mapsto \edges$. The columns of the discrete curl lie in the null space of the divergence, so that $\divgrid\curlgrid \equiv 0$. Because of the adjoint relationship between $\gradgrid$ and $\divgrid$, the discrete curl of the discrete gradient is also identically zero, $\rotgrid\gradgrid \equiv 0$. There is also a discrete gradient that acts upon the field of face-centered data and maps this to the tensor space: $\gradgrid_{\faces}: \faces \mapsto \tensors$. Its negative adjoint is a divergence operator $\divgrid_{\tensors}$ that maps tensor-valued data to vector-valued data, $\divgrid_{\tensors}: \tensors \mapsto \faces$. It is important to note that there is a discrete Laplacian for each grid space, and we use a subscript to denote this space, e.g., $\lapgrid_{\centers} \equiv \divgrid\gradgrid$ for the Laplacian acting on cell-centered data, and $\lapgrid_{\faces} \equiv \divgrid_{\tensors} \gradgrid_{\faces}$ for the Laplacian on face-centered data.

Grid data may be transformed on a staggered grid with second-order accuracy from one space to another via a simple two-point average, placing the result midway between the points. For example, we denote the transformation of cell-centered data to face data by $\interpgrid{\centers}{\faces}$. This operation forms $d$ vector components at each respective face of cell $\idex$ by averaging the two cell-centered values on either side of the face. In fact, this interpolation invokes the same pairs of points as the discrete gradient, but with differences replaced by averages. The opposite transformation, denoted by $\interpgrid{\faces}{\centers}$, is the adjoint of $\interpgrid{\centers}{\faces}$:
\begin{equation}
\ipcenters {\gridvar{\solnvec}}{\interpgrid{\faces}{\centers}\gridvar{\vecvec}} = \ipfaces {\interpgrid{\centers}{\faces}\gridvar{\solnvec}}{\gridvar{\vecvec}},
\end{equation} 
for $\gridvar{\solnvec} \in \centers$, $\gridvar{\vecvec} \in \faces$. The transformation $\interpgrid{\faces}{\centers}$ takes the average of the values from each of the two faces adjacent to a cell center oriented in a direction and adds all $d$ such averages together. As such, it is the additive companion to the discrete divergence.

Another useful interpolation, $\interpgrid{\faces}{\tensors}$, maps vector-valued elements in $\faces$ to tensor-valued elements in $\tensors$, and acts as a companion to $\gradgrid_{\faces}$. This operation computes an average of each of the $d$ face-centered components on a cell in $d$ orthogonal directions; the colored lines that intersect the face-centered points in Figure~\ref{fig:unitcell} indicate the directions in which the averaging is carried out. For example, for a vector $\gridvar{\vecvec} \in \faces$, the $x$ component $\gridvar{\vecvec}^{(1)}$ is averaged in the $x$ direction to the $\centers$ space, in the $y$ direction to the $\edges_{z}$ space, and the $z$ direction to the $\edges_{y}$ space. If we view the components of the resulting tensor as a $d\times d$ matrix, as in (\ref{eq:tensors}), these interpolations of $\gridvar{\vecvec}^{(1)}$ would fill the first column of this tensor. The adjoint operator, $\interpgrid{\tensors}{\faces}$, performs an interpolation of the entries in each of the $d$ columns in the respective directions of the $d$ rows and then adds the interpolated entries in the rows together to generate a component of an element of $\faces$; it is a companion to the $\divgrid_{\tensors}$ operator.

The interpolation transformations are important for enabling products of grid vectors from different spaces, particularly when these products are acted upon by the differencing operators: in such cases, one seeks to preserve a discrete analog of the product rule with second-order accuracy. Two particularly useful discrete product rules we will need later are
\begin{equation}
\label{eq:gradprod}
\gradgrid(\gridvar{\solnvec}_{1}\had\gridvar{\solnvec}_{2}) = \gradgrid\gridvar{\solnvec}_{1} \had \interpgrid{\centers}{\faces} \gridvar{\solnvec}_{2} + \interpgrid{\centers}{\faces} \gridvar{\solnvec}_{1} \had \gradgrid\gridvar{\solnvec}_{2},
\end{equation}
where $\gridvar{\solnvec}_{1}, \gridvar{\solnvec}_{2} \in \centers$, which approximates the gradient of a product of scalar fields to second-order accuracy; and
\begin{equation}
\label{eq:divprod}
\divgrid(\interpgrid{\centers}{\faces}\gridvar{\solnvec}\had \gridvar{\vecvec}) = \interpgrid{\faces}{\centers}\left(  \gridvar{\vecvec} \had \gradgrid \gridvar{\solnvec}\right) + \gridvar{\solnvec} \had \divgrid \gridvar{\vecvec},
\end{equation}
for $\gridvar{\solnvec} \in \centers$ and $\gridvar{\vecvec} \in \faces$, which approximates the divergence of a scalar--vector product. The first term on the right-hand side of this latter identity represents the dot product between the gradient of the scalar field and the vector field; the transformation $\interpgrid{\faces}{\centers}$ in this term is responsible for adding the $d$ components in the dot product at each grid location.

Furthermore, using the interpolations between $\faces$ and $\tensors$, we can also define a tensor product, $\tensprod$, between any two vector-valued elements $\gridvar{\vecvec}, \gridvar{v} \in \faces$:
\begin{equation}
\gridvar{\vecvec} \tensprod \gridvar{v} \equiv \left(\interpgrid{\faces}{\tensors} \gridvar{\vecvec}\right)^{T} \had \interpgrid{\faces}{\tensors} \gridvar{v} \in \tensors,
\end{equation}
where the $T$ superscript indicates that the resulting tensor in each cell should be transposed. With this tensor product defined, we can write a product rule identity analogous to (\ref{eq:divprod}) for the divergence of tensor products of two vector fields,
\begin{equation}
\label{eq:discretedivuv}
\divgrid_{\tensors} \left( \gridvar{\vecvec} \tensprod \gridvar{v} \right) = \interpgrid{\tensors}{\faces} \left( \left(\interpgrid{\faces}{\tensors} \gridvar{\vecvec}\right)^{T}\had \gradgrid_{\faces} \gridvar{v}\right)  + \interpgrid{\centers}{\faces} \divgrid \gridvar{\vecvec} \had \gridvar{v}.
\end{equation}
This is the discrete analog of the identity $\div(\vecfunc\boldsymbol{v}) = \vecfunc\cdot\grad\boldsymbol{v} + (\div \vecfunc)\boldsymbol{v}$. For a divergence-free $\gridvar{\vecvec}$, the left-hand side and the first term on the right-hand side are equivalent and either can be used for the convective term in Navier--Stokes.  

 In order to connect these grid spaces and their operators to the physical space $\R^{d}$, let us suppose that the cell center with index $\idex=0$ lies at position $\x_{\origin} \in \R^{d}$. On this uniform Cartesian grid, the position of any other cell center $\idex$ follows easily, $\xcenter{\idex} = \idex\dx + \x_{\origin}$. The points in all other grid spaces have similar coordinates, shifted by half a cell in appropriate directions as indicated by Figure~\ref{fig:unitcell}. For example, the $x$ coordinates in $\faces_{x}$ are described by $\xxface{x}{\idex} = (i_{1}+1/2)\dx + x_{\origin}$, the $y$ coordinates in $\faces_{y}$ by $\xyface{y}{\idex} = (i_{2}+1/2)\dx + y_{\origin}$, and the $z$ coordinates in $\faces_{z}$ follow in the obvious manner. We will collectively refer to these as $\xface{\idex}$. The coordinates $\xedge{\idex}$ of the edge space follow by inspection of the unit cell in Figure~\ref{fig:unitcell}.

\subsubsection{The lattice Green's function and relevant properties}

The lattice Green's function $\lgf_{0}$ is the solution of the discrete analog of equation~(\ref{eq:green}), namely, the algebraic set of equations
\begin{equation}
\label{eq:lgf}
\lapgrid \lgf_{0} = -\unitgrid_{0},
\end{equation}
where $\lapgrid$ is the discrete Laplacian operator belonging to one of the grid spaces. Analogous to the continuous Green's function, $G$, the lattice Green's function is invariant to translation of the source. That is, if we change the right-hand side of (\ref{eq:lgf}) to $-\unitgrid_{\idex_{0}}$, then the solution is a vector $\lgf_{\idex_{0}}$ whose components are $\gridcomp{\lgfcomp_{\idex_{0}}}{\idex} = \gridcomp{\lgfcomp_{0}}{\idex-\idex_{0}}$.

Consider a general discrete Poisson problem on the same set of points,
\begin{equation}
\label{eq:gridpoisson}
\lapgrid \gridvar{\solnvec}  = \gridvar{\rhsvec},
\end{equation}
where $\gridvar{\solnvec}$ and $\gridvar{\rhsvec}$ belong to the same grid space upon which $\lapgrid$ acts (e.g., $\centers$ or $\faces$). We can always decompose the right-hand side $\gridvar{\rhsvec}$ into grid unit vectors as in equation~(\ref{eq:celldecomp}); substituting for these unit vectors with (\ref{eq:lgf}), we have
\begin{equation}
\lapgrid\gridvar{\solnvec} = \sum_{\idex} \gridcomp{\rhsvec}{\idex}\unitgrid_{\idex}  = -\sum_{\idex} \gridcomp{\rhsvec}{\idex} \lapgrid \lgf_{\idex} = -\lapgrid \sum_{\idex} \gridcomp{\rhsvec}{\idex} \lgf_{\idex}.
\end{equation}
This suggests that the inhomogeneous part of the solution of (\ref{eq:gridpoisson}) is
\begin{equation}
\gridvar{\solnvec} = -\sum_{\idex}\gridcomp{\rhsvec}{\idex} \lgf_{\idex}.
\end{equation}
We can thus form the inverse of $\lapgrid$ by constituting its columns with the vectors $-\lgf_{\idex}$, where $\idex$ corresponds to the column. With the inverse $\invlapgrid$ so formed, the particular solution of (\ref{eq:gridpoisson}) can be written (for vectors in any grid space)
\begin{equation}
\gridvar{\solnvec} = \invlapgrid\gridvar{\rhsvec}.
\label{eq:lapgrid}
\end{equation}

In (\ref{eq:divlap}) we note that the divergence and the Laplacian commute with one another. We note that a similar identity can be shown for the interpolation $\interpgrid{\faces}{\centers}$:
\begin{equation}
\interpgrid{\faces}{\centers} \lapgrid_{\faces} = \lapgrid_{\centers} \interpgrid{\faces}{\centers}.
\end{equation}

\subsection{Spaces and operators of immersed points} Let us consider a point $\x \in \R^{d}$ that does not necessarily coincide with one of the grid points. We enlist the help of the DDF to {\em immerse} the point and its associated data into the grid. For example, a value $F$ associated with this immersed point is transferred to the cell centers by
\begin{equation}
F \ddf(\xcenter{\idex} - \x),
\end{equation}
for all multi-indices $\idex$.

The DDF has the form of a Cartesian product over one-dimensional functions,
\begin{equation}
\ddf(\x) = \frac{1}{\dx^{d}} \phi(x_{1}/\dx)\phi(x_{2}/\dx)\cdots \phi(x_{d}/\dx),
\end{equation}
where $\phi(r)$ is a continuous function satisfying various properties to ensure that $\ddf$ mimics the behavior of the actual Dirac delta function, $\dirac$, but on the grid rather than in $\R^{d}$ \citep{roma99}. Most importantly, $\ddf$ obeys the discrete analog of the integral constraint (\ref{eq:siftdiracd}),
\begin{equation}
\label{eq:sumddf}
\dx^{d}\sum_{\idex} \ddf(\x - \idex\dx) = 1,
\end{equation} 
for any $\x \in \R^{d}$. The sum in this identity is carried out over all grid points, but in practice, the DDF generally has compact support, ensuring that only grid points in a small neighborhood of $\x$ are invoked in the summation. In the current work, we favor the smoothed 3-point DDF proposed by Yang et al.~\cite{yang2009smoothing}, but the results are similar for other choices.

A collection of $\numpts$ points, with positions $\X_{\pindex}$, $\pindex = 1,\ldots,\numpts$, is immersed into the grid simply by a linear superposition of the points in the collection. This immersion process is succinctly represented by a linear {\em regularization operator}, $\reg$, for which each of the $\numpts$ columns corresponds to a particular immersed point, each row to a specific grid point, and each entry to the value of the DDF evaluated at the difference between these two points. On a staggered grid, the physical locations of each grid space are different, so the entries in the regularization operator depend on the grid space into which the data is mapped. For example, for regularizing scalar-valued data associated with the immersed points (a space which we will denote by $\spoints$) to cell centers on the grid, we define $\reg_{\centers}: \spoints \mapsto \centers$, for which a typical entry would be $\ddf(\xcenter{\idex} - \X_{\pindex})$. Similarly, $\reg_{\faces}: \vpoints \mapsto \faces$ regularizes vector-valued immersed point data (a space we denote by $\vpoints$) onto the cell faces, and $\reg_{\tensors}: \tpoints \mapsto \tensors$ regularizes tensor-valued point data (in $\tpoints$) to the cell tensor space. In some cases, one has need to regularize vector point data to cell edges, in which case one would define $\reg_{\edges}: \vpoints \mapsto \edges$. Each of the components of these vectors and tensors is regularized independently to the corresponding points on the grid.

The regularization operation does not rely on any particular relationship between the immersed points in the collection; they are each treated independently by the DDF. Indeed, this feature is generally regarded as an advantage of immersed boundary methods based on the DDF \cite{peskin:1j,roma99,taira2007,liskacolonius17}, since it allows one to enforce boundary conditions on a continuous surface by simply sampling the surface with points $\X_{\pindex}$, $\pindex = 1,\ldots,N$, rather than subdividing it into connected elements. However, in order to develop the desired generalized operators in this work, we will need to supply surface information: namely, we will need to assign a surface area $\dS_{\pindex}$ and unit normal vector $\nrm_{\pindex}$ to each immersed point $\pindex$, each approximating the local characteristics of the surface $\indic=0$ at the corresponding point $\X_{\pindex}$ to some degree. Later in this section we will propose details for the discrete versions of these surface characteristics; for now, we will simply denote the vector of surface areas by $\dSvec \in \spoints$ and vector of normals by $\normvec \in \vpoints$. We will also utilize the vector of assembled point coordinates $\X_{\pindex}$, as well, and denote this by $\xpointvec \in \vpoints$.

For notational purposes, it is useful to define some operations between vectors in $\spoints$ and $\vpoints$. An element-by-element product of two vectors in $\spoint{a}, \spoint{b} \in \spoints$ will be denoted simply by $\spoint{c} = \spoint{a}\had\spoint{b} \in \spoints$. The same notation will be used for an element-by-element product of $\spoint{a} \in \spoints$ with $\vpoint{v} \in \vpoints$, with the understanding that each ``element'' of $\vpoint{v}$ comprises the $d$ components of a vector in $\R^{d}$; the resulting product $\spoint{a}\had\vpoint{v}$ consists of each such $d$-dimensional element of $\vpoint{v}$ multiplied by the corresponding scalar element of $\spoint{a}$ and represents a vector in $\vpoints$. We can also define other standard vector operations between two vectors $\vpoint{u},\vpoint{v} \in \vpoints$ that act in the straightforward way on corresponding elements in the vectors, such as cross product $\vpoint{u}\times\vpoint{v} \in \vpoints$, dot product $\vpoint{u}\cdot\vpoint{v} \in \spoints$, and tensor product $\vpoint{u}\tensprod\vpoint{v} = \tpoint{T} \in \tpoints$. In this latter operation, the components of the resulting tensor at each immersed point $\pindex$ are
\begin{equation}
T_{ij\pindex} = u_{i\pindex}v_{j\pindex}.
\end{equation}

We can define inner products on both $\spoints$ and $\vpoints$ that approximate a continuous surface integral by weighting each element product by the corresponding surface area $\dS_{\pindex}$; for example, for any two vectors $\spoint{a}_{1},\spoint{a}_{2} \in \spoints$, with respective components $a_{1\pindex}$ and $a_{2\pindex}$, the inner product on $\spoints$ is
\begin{equation}
\ipscalar{\spoint{a}_{1}}{\spoint{a}_{2}} = \sum_{\pindex=1}^{N} a_{1\pindex} a_{2\pindex} \dS_{p}.
\end{equation}
As a special case, is we let $\onesspoint \in \spoints$ denote the special vector uniformly equal to 1 at all immersed points, then
\begin{equation}
\ipscalar{\onesspoint}{\spoint{a}} = \sum_{\pindex=1}^{N} a_{\pindex} \dS_{p}
\end{equation}
approximates a surface integral of function $a(\surfc)$, sampled on the immersed points $\X_{\pindex}$, $\pindex = 1,\ldots,N$. The squared norm of $\onesspoint$ under this inner product, $\normscalar{\onesspoint}^{2} = \ipscalar{\onesspoint}{\onesspoint}$, is equal to the area of the discretized surface. 

The inner product over $\vpoints$ must sum the $d$ components of each vector-valued element. For $\vpoint{v}_{1},\vpoint{v}_{2} \in \vpoints$
\begin{equation}
\ipvector{\vpoint{v}_{1}}{\vpoint{v}_{2}} = \sum_{k=1}^{d}\sum_{\pindex=1}^{N} v^{(k)}_{1\pindex} v^{(k)}_{2\pindex} \dS_{\pindex}.
\end{equation}
We let, e.g., $\onesspoint^{(2)} \in \vpoints$ denote the vector with all entries equal to 1 in the $y$ direction and equal to 0 in the other $d-1$ direction(s). For example, the discrete version of the basic surface integral identity (\ref{eq:normint}) is
\begin{equation}
\label{eq:normident}
\ipvector{\onesspoint^{(k)}}{\normvec} = 0,
\end{equation}
for $k = 1,\ldots,d$. We should insist on this discrete identity as a constraint for establishing the discretized surface normals. Indeed, it is generally straightforward to satisfy the constraint by defining the normals from differencing of the points. For example, in two dimensions, we can define the normal of any point $\pindex$ as
\begin{equation}
\label{eq:normal2d}
\boldsymbol{n}_{\pindex} = (\delta Y_{\pindex}/\dS_{\pindex}, -\delta X_{\pindex}/\dS_{\pindex}),
\end{equation}
where $\delta X_{\pindex} = \frac{1}{2} (X_{\pindex+1} - X_{\pindex-1})$, $\delta Y_{\pindex} = \frac{1}{2} (Y_{\pindex+1} - Y_{\pindex-1})$, and $\dS_{\pindex} = (\delta X_{\pindex}^{2}+\delta Y_{\pindex}^{2})^{1/2}$. In three dimensions, one can define the normals from cross products of edge vectors along triangular surface elements.

\subsubsection{Discrete analogs of integral identities.} With the surface information in $\dSvec$ and $\normvec$, we can now immediately construct discrete analogs of some of the continuous identities in \ref{sec:genindic}. For example, for scalar-valued data $\spoint{\solnvec}\in\spoints$, the discrete analog of the immersion operation~(\ref{eq:diracg3}) is $\reg_{\centers} (\dSvec\had\spoint{\solnvec})$, where we have enclosed the element-by-element product of $\dSvec$ and $\spoint{\solnvec}$ in parentheses to clarify that this operation precedes the regularization. This operation on immersed point data is so important for our later results that we define it as the {\em area-weighted regularization operator}, $\regds$. For regularizing scalar data to cell centers, vector data to cell faces or edges, and tensor data to cell tensor locations, we define, respectively,
\begin{align}
\label{eq:ddfg2}
\regds_{\centers} \equiv  \reg_{\centers}(\dSvec\had\cdot) : \spoints \mapsto \centers,&\qquad \regds_{\faces} \equiv  \reg_{\faces}(\dSvec\had\cdot) : \vpoints \mapsto \faces,\\
\regds_{\edges} \equiv  \reg_{\edges}(\dSvec\had\cdot) : \vpoints \mapsto \edges,&\qquad\regds_{\tensors} \equiv  \reg_{\tensors}(\dSvec\had\cdot) : \tpoints \mapsto \tensors \nonumber.
\end{align}
In fact, these definitions for $\regds$ serve as the discrete analog of (\ref{eq:diracg2unity}), and thus, \emph{$\regds$ represents the discrete version of the immersion function $\dirac(\indic)$}. Like $\dirac(\indic)$, $\regds$ has units of inverse length; in the discrete context, this length represents the breadth of the region surrounding the points over which immersed data is spread.

Let us consider the inner product of some scalar grid data $\gridvar{\solnvec} \in \centers$ with immersed point data, $\spoint{\scavec}\in\spoints$, regularized to the cell centers. Using their definitions, it is easy to show that the first inner product over $\centers$ can be rewritten as an inner product over $\spoints$:
\begin{equation}
\label{eq:reginterp}
\ipcenters{\gridvar{\solnvec}}{\regds_{\centers}\spoint{\scavec}} = \ipscalar{\dx^{d}\reg_{\centers}^{T}\gridvar{\solnvec}}{\spoint{\scavec}} = \ipscalar{\interp_{\centers}\gridvar{\solnvec}}{\spoint{\scavec}},
\end{equation}
where $()^{T}$ denotes the matrix transpose. In the last equality, we have defined a new operator, $\interp_{\centers}$, called the {\em interpolation operator} \citep{roma99}, which transfers grid cell-centered data to the immersed points,
\begin{equation}
\interp_{\centers} \equiv \dx^{d}\reg_{\centers}^{T} : \centers \mapsto \spoints;
\end{equation}
analogous interpolation operators, $\interp_{\faces}$ and $\interp_{\tensors}$, can be readily defined for transferring cell-face data in $\faces$ or cell-edge data in $\edges$ to $\vpoints$ and cell-tensor data in $\tensors$ to $\tpoints$. Interpolation represents the discrete form of the restriction function $\dirac^{T}(\indic)$ (\ref{eq:restrict}) and has no underlying units. And just as the immersion and restriction operators are transposes of one another, {\em interpolation is the adjoint of area-weighted regularization} with respect to the inner products defined in this paper. Since it is formed from the transpose of $\reg_{\centers}$, this interpolation operator acts only on grid data within the support of the DDF centered at the immersed points.

It is important to note that the property (\ref{eq:sumddf}) ensures that
\begin{equation}
\interp_{\centers}\onesgrid = \onesspoint;
\end{equation}
a similar result holds in each direction for the other interpolation operators. In other words, the interpolation operator preserves the value of uniform data on the grid. By setting $\gridvar{\solnvec} = \onesgrid$ in equation~(\ref{eq:reginterp}) and switching the sides of the inner products, this property of the interpolation operator $\interp$ ensures, in turn, that
\begin{equation}
\label{eq:ipscalarcenter}
\ipscalar{\onesspoint}{\spoint{u}} = \ipcenters{\onesgrid}{\regds_{\centers}\spoint{u}},
\end{equation}
or, in words, that the area-weighted sum of immersed point data $\spoint{u}$ is preserved by the volume-weighted sum of the data after it is regularized to the grid. This is the discrete analog of equation~(\ref{eq:diracg}).


\section*{References}

\bibliographystyle{plain}

\end{document}